\documentclass{aa}  

\usepackage{graphicx}
\usepackage{txfonts}
\def\kms{km~s$^{-1}$}

\usepackage{natbib}
\bibpunct{(}{)}{;}{a}{}{,}
\usepackage{supertabular}
\usepackage{rotating}
\usepackage{setspace}
\usepackage{lscape}
\usepackage{color}

\begin{document} 


   \title{Luminous blue variables: An imaging perspective on their binarity and near environment\thanks{Based on observations collected at the European Organisation for Astronomical Research in the Southern Hemisphere, Chile, under projects number 085.D-0625(C), 087.D-0426(C,D), and archival data 383.D-0323(A).}}

   \titlerunning{LBVs: imaging and binarity}

\author{
Christophe Martayan \inst{1}
\and Alex Lobel \inst{2}
\and Dietrich Baade \inst{3}
\and Andrea Mehner \inst{1}
\and Thomas Rivinius \inst{1}
\and Henri M. J. Boffin \inst{1}
\and Julien Girard \inst{1}
\and Dimitri Mawet \inst{4}
\and Guillaume Montagnier \inst{5}
\and Ronny Blomme \inst{2}
\and Pierre Kervella \inst{6,7}
\and Hugues Sana \inst{8}
\and Stanislav \v{S}tefl\thanks{Deceased} \inst{9}
\and Juan Zorec  \inst{10}
\and Sylvestre Lacour \inst{6}
\and Jean-Baptiste Le Bouquin \inst{11}
\and Fabrice Martins \inst{12}
\and Antoine M\'erand \inst{1}
\and Fabien Patru \inst{11}
\and Fernando Selman \inst{1}
\and Yves Fr\'emat \inst{2}
}

\institute{European Organisation for Astronomical Research in the Southern 
Hemisphere, Alonso de C\'ordova 3107, Vitacura, Casilla 19001, Santiago de Chile, Chile
\email{Christophe.Martayan@eso.org}
\and Royal Observatory of Belgium, 3 Avenue Circulaire, 1180 Brussels, Belgium 
\and European Organisation for Astronomical Research in the Southern 
Hemisphere, Karl-Schwarzschild-Str.\ 2, 85748 Garching b.\ M\"unchen, Germany
\and Department of Astronomy, California Institute of Technology, 1200 E. California Blvd, MC 249-17, Pasadena, CA 91125 USA
\and Observatoire de Haute-Provence, CNRS/OAMP, 04870 Saint-Michel-l'Observatoire, France 
\and LESIA, UMR 8109, Observatoire de Paris, CNRS, UPMC, Univ. Paris-Diderot, PSL, 5 place Jules Janssen, 92195 Meudon, France
\and Unidad Mixta Internacional Franco-Chilena de Astronom\'{i}a (UMI 3386), CNRS/INSU, France \& Departamento de Astronom\'{i}a, Universidad de Chile, Camino El Observatorio 1515, Las Condes, Santiago, Chile
\and ESA / Space Telescope Science Institute, 3700 San Martin Drive, Baltimore, MD 21218, United States of America
\and ESO/ALMA - The Atacama Large Millimeter/Submillimeter Array, Alonso de C\'ordova 3107, Vitacura, Casilla 763 0355, Santiago, Chile
\and Sorbonne Universit\'es, UPMC Universit\'e Paris 6 et CNRS, UMR7095 Institut d'Astrophysique de Paris, F-75014 Paris, France
\and UJF-Grenoble 1/CNRS-INSU, Institut de Plan\'etologie et d'Astrophysique de Grenoble (IPAG) UMR 5274, BP 53, 38041 Grenoble C\'edex 9, France
\and LUPM, Universit\'e de Montpellier, CNRS, Place Eug\`ene Bataillon, F-34095 Montpellier Cedex 05
}

   \date{Received; accepted}

 
  \abstract
   {Luminous blue variables (LBVs) are rare massive stars with very high luminosity. They are characterized by strong photometric and spectroscopic variability related to transient eruptions.
The mechanisms at the origin of these eruptions is not well known. In addition, their formation is still problematic and the presence of a companion could help to explain how they form.}
   {This article presents a study of seven LBVs (about 20\% of the known Galactic population),  some Wolf-Rayet stars, and massive binaries.  We probe the environments that surround these massive stars with near-, mid-, and far-infrared
   images, investigating potential nebula/shells and the companion stars.}
   {To investigate large spatial scales, we used seeing-limited and near diffraction-limited adaptive optics images to obtain a differential diagnostic on the presence of circumstellar matter and
to determine their extent. From those images, we also looked for the presence of binary companions on a wide orbit. Once a companion was detected, its gravitational binding to the central star was tested.
 Tests include the chance projection probability, the proper motion estimates with multi-epoch observations, flux ratio, and star separations. }
   {We find that two out of seven of LBVs may have a wide orbit companion. Most of the LBVs display a large circumstellar envelope or several shells. In particular, HD168625, known for its rings, possesses several
shells with possibly a large cold shell at the edge of which the rings are formed.  For the first time, we have directly imaged the companion of LBV stars.
}
 {}


   \maketitle
%

\section{Introduction}

Luminous blue variables (LBVs) are rare massive stars that range in size from  tens to more than 100 M$_{\odot}$. Their luminosity can reach or exceed the Humphreys-Davidson limit \citep{hd1979} with log$(L/L_{\odot})$=5 to 7. They are surrounded by massive envelopes created during major episodic eruptions with 0.1 to 20 M$_{\odot}$ ejected and mass loss rates of 10$^{-4}$ to 10$^{-5}$ M$_{\odot}$ yr$^{-1}$ \citep{HD1994}  suggesting super-Eddington winds \citep{vm2008} in active phase. 
They have mass loss rates of 10$^{-7}$ to 10$^{-6}$ M$_{\odot}$ yr$^{-1}$ during quiescence. 

Luminous blue variables show luminosity variations with different time-scales from days to decades \citep{van2001}. 
LBVs that  show \object{S Dor} phase are fast rotators with velocities reaching the critical regime in their minimum phase \citep{groh2009} and according to theoretical models, it is possible that they  have strong polar winds \citep{md2001,owocki2011}. Their surrounding nebulae are often axisymmetric, such as the Homunculus nebula around \object{$\eta$ Car}. 
However, \cite{soker2004} argues that a single-star model cannot reproduce these bipolar nebulae.

Interestingly, several massive stars exhibit nebular rings.
One proposed way to form such rings is through the merger of a binary. The outer rings of \object{SN 1987A} in particular display a likeness with those around stars such as HD168625 and \object{Sher 25} \citep{morris2009}. This has led some to suggest that these stars may be good candidates for Galactic supernovae \citep[][ and references therein]{smith2007}.


The cause of the major LBV eruptions is still debated. Different types of instabilities are proposed to explain them, among which radiation and turbulent pressure instabilities, vibrations, and dynamical instabilities \citep{HD1994}. The constant dense slow wind of LBVs could play a role as well \citep{HD2014}. Binarity with mass transfer \citep{kashi2010,kashi2010b} is also a promising explanation.

During their giant eruptive phase, extragalactic LBVs in the early eruption stages can be confused
with normal supernovae (SNe). These events are sometimes called supernova impostors \citep{vd2012}. According to theory, LBVs should not explode as core-collapse SNe without passing through the Wolf-Rayet (WR) phase.
However, recent SN radio lightcurve observations tend to demonstrate the opposite. They show the SN ejecta interacting with large amounts of circumstellar matter. It appears to be related to strong mass-loss episodes that occurred before the SN. LBVs during their S Dor phase with their giant outbursts would best explain them \citep{kotak2006}.
Other observations support this scenario, spectroscopically in \object{SN 2005gj} \citep{trundle2008}, and with direct identification of the progenitor in images in the cases of \object{SN 2005gl} \citep{galyam2009} and \object{SN 2010jl} \citep{smith2011}.
Moreover, \cite{groh2013} recently found that models can explain some SNe IIb by direct explosion of a relatively low-mass LBV star.
This kind of LBV evolution could lead to the potential ultra-powerful pair-instability supernova \citep[for the more massive of them with masses above 100 M$_{\odot}$, see][]{fryer2001}.

\cite{smith2015} argue that because the LBVs are mostly isolated they probably result from binary evolution and can be considered  massive evolved blue stragglers. This would imply that various types of massive stars could eventually become LBVs through binary mergers.

At the other end of  stellar evolution, the formation is still not well understood, although models involving relatively distant low-mass companions  \citep{krum2012} may offer a decisive clue. Binarity may play a critical role in the formation of very massive objects with some protostellar collisions \citep{baum2011,tan2014}. Moreover, it is expected that  fragmentation could lead to the presence of lower mass companions with separations of hundreds or thousands of au \citep{krum2009,krum2012}.
  
However, to date very few LBV stars are known with certainty to be binaries: in the Galaxy there is $\eta$ Car \citep{dam1997}, MWC314 \citep{lobel2013}, and HR Car \citep{rivinius2014}. In addition, LBV1806-20 was found to be a double-lined binary by \cite{figer2004}, but it still lacks a spectroscopic follow-up to confirm its binary nature and to discard a contamination by another star in this crowded field.
With VLTI-AMBER \citep{amber} observations and with radial velocity variations, and the line shape modifications found in X-Shooter \citep{xshooter} spectra, \cite{martayan2012} indicate that the Pistol Star could also be binary.
In other galaxies only \object{HD5980}  is known as a binary \citep{foellmi2008,georgiev2011}. 
 
It is therefore important to investigate in a deeper and homogeneous way the presence of potential companions around LBV stars. So far, there has been no systematic search for companions of LBVs. 

With very large telescopes, seeing-limited imaging probes scales $\sim$1$\arcsec$, while with adaptive optics (AO) imaging the limit decreases to 60 mas \citep{sanascan2010}. 
In this article, we investigate the presence of potential companions in wide orbits with various imaging techniques mainly at IR wavelengths. Future investigations will focus on smaller spatial scales to find potential companions in closer orbits. Preliminary results can be found in \cite{martayan2012}.
The presence of a companion provides constraints on models of star formation and evolution. Furthermore, it also gives clues to the shaping of the surrounding nebulae and possibly to the triggering of giant LBV eruptions.

In Section~\ref{obs}, we present our observational and archival data.
Section~\ref{binarity} deals with the methods (chance projection probability, multi-epoch proper motion, etc.) used to determine whether the nearby objects are bound to the main stars, and discusses the results. The presence of the shells around the stars of the sample is reported in Section ~\ref{GCED} with a detailed example from HD168625.
Section~\ref{conclue} provides a summary. In Appendix~\ref{GCED2}, the images of the stars of the sample are displayed and their remarkable environment structures are discussed. In Appendix~\ref{catal}, small catalogues of stars in the NACO field of view of HD168625 and the Pistol Star are given.

\begin{table}[h] 
\caption[]{Stellar sample discussed in this paper, along with classification and coordinates from Simbad.
Candidate LBV stars are identified with ``cLBV''. They are ordered following their classification and right ascension.}
\tiny{
\begin{tabular}{@{\ }l@{\ \ }l@{\ \ }l@{\ \ }l@{\ \ }l@{\ \ }l@{\ \ }l@{\ \ }}
\hline 
Star & Classification & Distance & RA(2000) & DEC(2000)  \\
 &  & kpc & h mn s & $\degr$ $\arcmin$ $\arcsec$ \\
\hline           
\object{Pistol Star} & cLBV               & 8$^{1}$ & 17 46 15.24 & -28 50 03.58 \\
\object{WR102ka} & LBV, WN10              & 8$^{1}$ & 17 46 18.12 & -29 01 36.60  \\
\object{LBV1806-20} & cLBV                & 11.8$^{2}$ & 18 08 40.31 & -20 24 41.10  \\
\object{HD168625} & cLBV, B6Ia            & 2.2$^{3}$ (2.8$^{4}$) & 18 21 19.55 & -16 22 26.06  \\
\object{HD168607} & LBV, B9Ia             & 2.2$^{3}$ & 18 21 14.89 & -16 22 31.76  \\
\object{MWC930} & LBV, B5-B9e             & 3.5$^{5}$  & 18 26 25.24 & -07 13 17.80  \\
\object{MWC314} & LBV, B3Ibe              & 3$^{6}$ &   19 21 33.98 & 14 52 56.89 \\
\hline
\object{HD152234} & O9.7Ia+O8V$^{a}$      & 1.91$^{7}$ & 16 54 01.84 & -41 48 23.01  \\
\object{WR102e} & WR, WC, ``dustar''$^{b}$& 8$^{1}$ & 17 46 14.81 & -28 50 00.60 \\
\object{HD164794} & O3.5Vf$^{+}$+O5Vf$^{c}$    & 1.1-1.8$^{8,9,10}$ & 18 03 52.45 & -24 21 38.63  \\
\hline
 \end{tabular}
\label{starsample}
\begin{flushleft}
a: Classification from \cite{sana2008}.\\
b: Star is in the Pistol Star field, and is considered to be a WR-``dustar'' by \cite{marchenko2007}.\\
c: Classification from \cite{rauw2012}.\\
1: \cite{reid1993}, 2: \cite{figer2004}, 3: \cite{chentsov2004}, 4: \cite{hut1994}, \cite{pasquali2002}, 5: \cite{miro2005}, 6: \cite{miro1998}, 7: \cite{ankay2001}, 8: 1.58 kpc,\cite{sung2000}, 9: 1.79 kpc,\cite[][ and references therein]{blomme2014}, 10: 1.1 kpc,\cite{bondar2012}.\\
\end{flushleft}
}
\end{table}

\section{Observations, archival data, and data reduction}
\label{obs}


The stars in our sample were selected for their properties to investigate the binary nature of LBVs and to study the envelope structure.
On the one hand the Pistol Star, WR102ka, and LBV1806-20 are stars located near the Humphreys-Davidson limit \citep{HD1994}, and on the other hand MWC314, MWC930, and WR102e are close to the limit where the rotational velocity reaches the critical velocity \citep{groh2009} for massive stars. 
HD168625 and HD168607 belong to the low-luminosity LBVs.
The massive binary stars HD164794 and HD152234 were added for comparison purposes.
The sample contains around 20\% of the Galactic LBV/candidate LBV stars (cLBVs) as defined by \cite{clark2005} and \cite{naze2012}. 
The sample is presented in Table~\ref{starsample} with information from Simbad.
Unfortunately, the parallax and proper motion values based on the Hipparcos or Tycho missions are not reliable for these very distant stars.

   
To study the binarity and the environment of the objects in Table~\ref{starsample}, two cameras with good spatial resolution were used.
The first one is NACO with its adaptive optics facility \citep{naco1,naco2} and the second is VISIR \citep{visir}. Both cameras allow near- to mid-IR observations.
Our observations are summarized in Table~\ref{sumobs}.

\begin{table}[t] 
\caption[]{Summary of the NACO and VISIR observations. The  service mode (SM) or visitor mode (VM) is indicated in col. 3. The length of the observations is given in hours or in nights in col. 4. Cols. 5 and 6 provide the weather condition: R-band image quality and sky transparency.}
\tiny{
\begin{tabular}{@{\ }l@{\ \ }l@{\ \ }l@{\ \ }l@{\ \ }l@{\ \ }l@{\ \ }l@{\ \ }}
\hline 
Date & Instrument & Mode & Length & Seeing & Transparency  \\
\hline           
17/06 + 01/08/2010 & NACO & SM & 4 $\times$ 1 h & 0.8$\arcsec$ & clear\\
17-18/07/2011 & NACO & VM & 1 n & 0.65$\arcsec$-0.8$\arcsec$ & clear\\
18-19/07/2011$^{1}$ & NACO & VM & 0.5 n &  0.55$\arcsec$-0.65$\arcsec$ & clear-thin \\
20/07/2011$^{2}$ & VISIR & VM & 0.5 n & 1.0$\arcsec$-1.5$\arcsec$& thin-thick \\
\hline           
 \end{tabular}
\label{sumobs}
\begin{flushleft}
1: a NACO technical problem prevented  observation of the star WR102ka.\\
2: a significant fraction of the planned observations were not performed owing to the weather.
\end{flushleft}
}
\end{table}

       \begin{figure*}[]
   \centering
    \includegraphics[width=15cm]{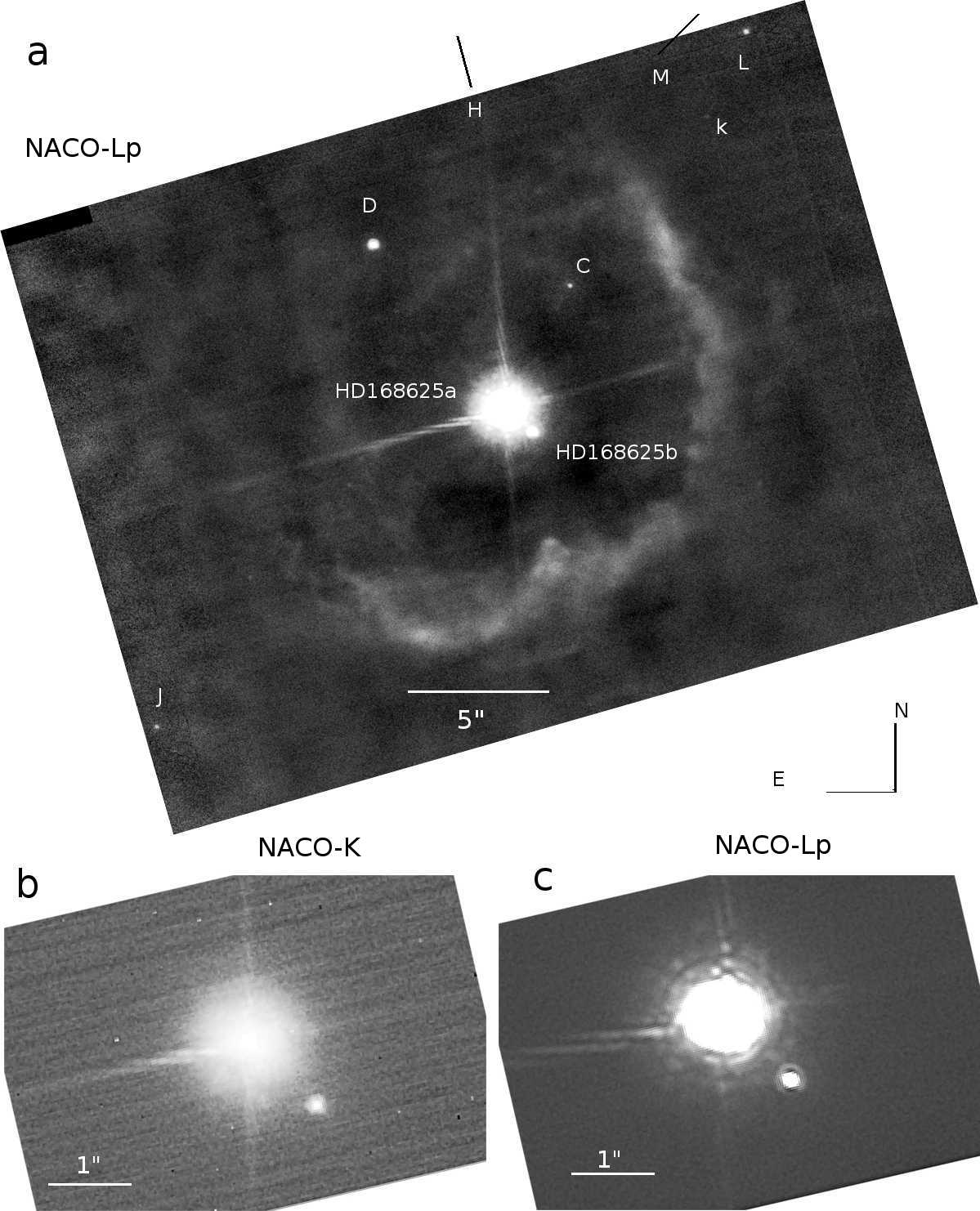}   
     \caption{Star HD168625, NACO images:
       a)  $28\arcsec \times 21\arcsec$ Lp image with labels of the surrounding stars; b):  $6\arcsec \times 4\arcsec$ zoom in K on the pair HD168625a,b.  The SW companion is at 1.15$\arcsec$ from the main star.
       c) $6\arcsec \times 4\arcsec$ zoom in Lp of HD168625a,b.}
         \label{Fighd168625b}
   \end{figure*}

The NACO observations were reduced with the ESO NACO pipeline version 4.3.3\footnote{ftp://ftp.eso.org/pub/dfs/pipelines/naco/naco-pipeline-manual-1.1.pdf} under EsoRex using the shift-and-add algorithm with dark and flat-field corrections. 
The VISIR observations were reduced using a new prototype pipeline version 4.0.0 under Reflex during beta-testing of the new interface \citep{reflex}. The images were also combined with a shift-and-add algorithm using the object detection in each image. If the object was not detected, offset values written in the fits header of the raw data files were used to make the image. In some cases, the covered field was not large enough to include the entire surrounding nebula (the Pistol
Star or HD168625). For the Pistol
Star our data did not have  enough signal to properly reconstruct the image (all observations of this target were stopped owing to windy observing conditions).  
Additional data were retrieved from ESO and other archives.
In particular, we used SINFONI \citep{sinfo1,sinfo2} data for WR102ka. We reduced the data using the Reflex interface and the pipeline version 2.5.2.1.
Moreover, we also used the GATOR database images from Spitzer \citep{spitzer} and its instruments IRAC \citep{irac} and MIPS \citep{mips}, and from WISE \citep{WISE}. Images from the MAST database, HST NICMOS \citep{nicmos}, and WFC \citep{wfc} and also  from other missions or instruments were obtained for various stars in order to perform multi-imaging techniques, multiwavelength and multi-epoch measurements.
The data from  Herschel \citep{herschell2010}, IUE \citep{iue1978}, SWIFT/UVOT \citep{swiftuvot2005}, and EMMI \citep{emmi1986} lack spatial resolution for detecting object proper motion. The summary of the images is available in Table~\ref{obssample} (incl. pixel sizes), while in Table~\ref{filtertable} the central wavelength of the filters used is mentioned.

\begin{table}[]
\centering
\caption[]{List of instruments used. The pixel size in angular units (mas or arcsec) of the cameras is indicated. 
The filters and NACO cameras (S13, S27, L27, L54) are mentioned. The central wavelength of each filter is provided in Table~\ref{filtertable}.
}
\tiny{
\begin{tabular}{@{\ }l@{\ \ }l@{\ \ }l@{\ \ }l@{\ \ }l@{\ \ }l@{\ \ }l@{\ \ }l@{\ \ }}
\hline 
Star & NACO & VISIR & Spitzer archive  \\
 & 13,27,54 mas & 75 mas & (irac 0.6$\arcsec$, MIPS 1.2$\arcsec$)  \\
\hline             
Pistol Star & K-S13/S27, Lp-L27 & PAH2*, SiC*, NeII* & irac1,2,3,4   \\
WR102ka & problem &  & irac1,2,3,4  \\
WR102ka & & & MIPS24    \\
LBV1806-20 & K-S27, Lp-L27 &  & irac1,2,3,4 \\
LBV1806-20 & &  &MIPS24   \\
HD168625 & 3.74/4.05/Lp-L54 & PAH1, PAH2 & irac1,2,3,4  \\
HD168625 & K-S27, Lp-L27 & SiC, NeII & MIPS24  \\
HD168607 & K-S27, Lp-L27 &  & irac1,2,3,4   \\
HD168607 &  &  & MIPS24   \\
MWC930 & K-S27, Lp-L27* &  &   \\
MWC314 & K-S27, Lp-L27 &  & irac1,2,3,4   \\
HD152234 &  & PAH1, SiC, NeII &    \\
WR102e & K-S13/S27, Lp-L27 &  & irac1,2,3,4   \\
HD164794 & K-S27, Lp-L27 &  & irac1,2,3,4   \\
HD164794 &  &  & MIPS24  \\
\hline
Star &  WISE-archive & HST-archive & other \\
 &  (1.38$\arcsec$) & 25,50,100 mas & archive \\
\hline
Pistol Star &1,2,3*,4* & yes & 2MASS** \\
WR102ka &  1,2,3,4 & no & SINFONI*** \\
LBV1806-20 &  1,2,3,4 & yes & N/A  \\
HD168625 &  1,2,3*,4 & yes & N/A \\
HD168607 &  1,2,3,4 & yes & N/A  \\
MWC930 &  1,2,3,4 & yes & N/A \\
MWC314 &  1,2,3,4 & no & swift*, EMMI*  \\
HD152234 &  1,2,3,4 & no & N/A \\
WR102e &  1,2,3*,4* & yes & N/A \\
HD164794 &  1,2,3,4 & yes* & VISIR* \\
\hline
\end{tabular}
\label{obssample}
\begin{flushleft}
*: too faint, saturated, not used (insufficient resolution or not deep enough).\\
**: \cite{2mass}.\\
***: SINFONI, scale 250 mas $\times$ 150 mas, data presented in \cite{os2013}.
\end{flushleft}
}
\end{table}

\begin{table}[]
\centering
\caption[]{Central wavelength in $\mu$m of filters used in this article.}
\tiny{
\begin{tabular}{@{\ }l@{\ \ }l@{\ \ }l@{\ \ }l@{\ \ }l@{\ \ }l@{\ \ }l@{\ \ }l@{\ \ }}
\hline 
Filter & $\lambda$ & Filter & $\lambda$ & Filter & $\lambda$ & Filter & $\lambda$\\
\hline 
2MASS-Ks & 2.15 & NACO-K & 2.19 & WISE1 & 3.4 & NACO-Lp & 3.45 \\
irac1 & 3.6 & irac2 & 4.5 &  WISE2 & 4.6 & irac3 & 5.8 \\
irac4 & 8.0 & VISIR-PAH1 & 8.59 & VISIR-PAH2 & 11.25 &  VISIR-SiC & 11.85  \\
WISE3 & 12 & VISIR-Ne2 & 12.81 & WISE4 & 22 & MIPS & 24 \\ 
\hline
\end{tabular}
\label{filtertable}
}
\end{table}

The images of HD168625 are shown in Figs.~\ref{Fighd168625b} and~\ref{Fighd168625IR},
while for the remaining objects of Table~\ref{starsample} the images are presented in the Appendix in Figs.~\ref{FigPS} to~\ref{Fighd164794}. They reveal the presence of nebular shell(s) around the main star. Close neighbour(s) are visible in the cases of the Pistol Star, HD168625, LBV1806-20, MWC314, and HD152234.
Basic PSF fitting was performed to obtain their coordinates. The detection cut-off was set at a minimum of 3-$\sigma$ over the background. The limiting magnitude depends on the instrument setup and exposure times and it reached 19-20 in K band and 13-15 in Lp. More than a limiting magnitude, the difference in magnitude (or the contrast at a given separation) could be more suited to characterizing the detection of a close companion.
They are reported in Table~\ref{tableprobacomp}.

\section{Binarity}
\label{binarity}

The binarity in massive stars is important because it might have an impact on
\begin{itemize}
\item the way they form. Recent models of massive star formation \citep{krum2009,krum2012,krum2015} indicate that the star is created in multiple systems with companions in different orbital scales (including wide ones of hundreds to tens of thousands of au);
\item the way they evolve \citep[mergers, matter exchange, close orbital scale,][]{baum2011,tan2014,krum2015};
\item the triggering of eruptions \citep[intermediate to nearby scale,][]{kashi2010,kashi2010b};
\item the shaping of the circumstellar environment (all orbital scales).
\end{itemize}

In this first article, we focus on wide-orbit companions that could provide some constraints on the star formation models and could help with constraints on the massive star formation process and on the shaping of the surrounding nebulae. 

The first step in determining whether a star is in a potential binary system is  detecting a potential companion.
To this aim, one needs to probe different spatial scales. The largest scale is checked with the imaging techniques, assisted or not with an adaptive optics facility. NACO observations were mostly used because this is the instrument with the best spatial resolution with pixels of 13, 27, and 50 mas. Some HST images were used as well.
The closest neighbours found to the main stars are listed in Table~\ref{tableprobacomp}. 

To determine the likelihood  that a potential companion is bound to the main star, we used  a combination of different techniques and criteria that are presented in the following sections.

\begin{table*}[]
\centering
\caption[]{
Separation, magnitude difference (left), chance projection probability (middle) depending on the sample and method used, and proper motion comment and binding status of potential companion to the main object (right).
Blue values are larger than the defined limit (see text).
Details of the columns are indicated at the bottom of the Table.
}
\tiny{
\begin{tabular}{@{\ }l@{\ \ }l@{\ \ }l@{\ \ }l@{\ \ }l@{\ \ }l@{\ \ }|l@{\ \ }l@{\ \ }l@{\ \ }l@{\ \ }l@{\ \ }|l@{\ \ }l@{\ \ }l@{\ \ }l@{\ \ }l@{\ \ }l@{\ \ }l@{\ \ }l@{\ \ }}
\hline 
Star   &  Proj. sep.  &  prob. sep.  &  prob. sep.$^{1}$  &  Max proj.  &  $\Delta$mag  &   P(field)$^{2}$    &  P(field)$^{2}$  & P$^{\prime}$(field)  &  P$^{\prime}$cut-off  &  P,P$^{\prime}$,P$^{\prime\prime}$(field)$^{3}$  &  Comments  &  Status\\
    & $\arcsec$ & $\arcsec$ &  au  &  sep. $\arcsec$ for  &  $\delta$mag$<$8.5$^{m}$  &   NACO &  trilegal\%  & trilegal  &   &  2MASS  &  Motion  &  \\
     &    &    &    &  10,000 au  &     &  \%  &  all,$\pm$2$^{m}$, $\pm$1$^{m}$  &  \%  &  \%  &  \%  &   &   \\
(1) & (2) & (3) & (4) & (5) & (6) & (7) & (8) & (9) & (10) & (11) & (12) & (13) \\
\hline                                                     
Pistol Star$\_{4}$   & 0.75 & 1.21 & 9698 & 0.77 &  3.1 - 2.3  &  \textcolor{blue}{11}  &  \textcolor{blue}{100}, 9, 5  & \textcolor{blue}{46}  & 42      & 5,4,5 &  A   &  borderline \\
Pistol Star$\_{3}$   & 1 & 1.61 &  \textcolor{blue}{12931}  & 0.77 &  4.1 - 3.7  &  \textcolor{blue}{20} &  \textcolor{blue}{100}, \textcolor{blue}{40}, \textcolor{blue}{15}  & \textcolor{blue}{66}  & 42      &  \textcolor{blue}{9,9,9}  &  \textcolor{blue}{B}   &  field \\
Pistol Star$\_{1}$   & 1.39 & 2.24 &  \textcolor{blue}{17974}  & 0.77 &  5 - 4.9  &  \textcolor{blue}{39} &  \textcolor{blue}{100}, \textcolor{blue}{100}, \textcolor{blue}{71}  & \textcolor{blue}{88}  & 42      &  \textcolor{blue}{16,15,16}  &  C  &  field\\
Pistol Star$\_{2}$   & 1.44 & 2.32 &  \textcolor{blue}{18620}  & 0.77 &  5 - 5  &  \textcolor{blue}{42} &  \textcolor{blue}{100}, \textcolor{blue}{100}, \textcolor{blue}{77}  & \textcolor{blue}{90}  & 42      &  \textcolor{blue}{17,15,15}  &  C   &  field \\
WR102ka$\_{b}$   & 1.35 & 1.49 &  \textcolor{blue}{11920}  & 0.77 &  $\sim$5  &   \textcolor{blue}{90} &  \textcolor{blue}{25}, 0, 0   & 22 & 42 &  \textcolor{blue}{13,12,13}  & D  &  field \\
WR102ka$\_{c}$   & 1.98 & 2.19 &  \textcolor{blue}{17520}  & 0.77 & 3.5 &  \textcolor{blue}{100}  &  \textcolor{blue}{52}, 0, 0  & 41 & 42 &  \textcolor{blue}{29,25,25}  & D  &  field \\
LBV1806-20b   & 0.76 & 1.23 &  \textcolor{blue}{14495}  & 0.52 & 6.6 &  \textcolor{blue}{20} &  5, 5, 4  & 18 & 62 & 3,3,3 &   &  field  \\
LBV1806-20c   & 0.9 & 1.45 &  \textcolor{blue}{17166}  & 0.52 & 6.6 &  \textcolor{blue}{28} &  8, 8, 5  & 24 & 62 & 4,4,4 &    &  field \\
LBV1806-20d  & 1.3 & 2.1 &  \textcolor{blue}{24795}  & 0.52 & 6.6 &  \textcolor{blue}{59} &  \textcolor{blue}{15}, \textcolor{blue}{15}, \textcolor{blue}{13}  & 44 & 62 &  \textcolor{blue}{9,9,9}  &     &  field \\
HD168625b   & 1.15 & 1.86 &  4089*  & 2.81 &  4.6 - 4.4  & 5 &  2, 1, 1  & $<$1  & 12      & 3,3,3 &  E   &  bound\\
HD168625b real$^{4}$  & 1.15 & 1.33 &  2927**  & 3.94 &  4.6 - 4.4  & 5 &  2, 1, 1  &  $<$1  & 12      & 3,3,3 &  E   &  bound\\
HD168607b  & 2.02 & 3.26 & 7183 & 2.81 &  \textcolor{blue}{10.1}  &  \textcolor{blue}{100} &  8, 7, 7  & \textcolor{blue}{65}  & 12      &   \textcolor{blue}{7,6,7}  &     &  field \\
HD168607c  & 4.19 & 6.76 &  \textcolor{blue}{14899}  & 2.81 &  \textcolor{blue}{9.1}  &  \textcolor{blue}{100}  &  \textcolor{blue}{31}, \textcolor{blue}{31}, \textcolor{blue}{16}  & \textcolor{blue}{99}  & 12      &  \textcolor{blue}{29,25,25}  &    &  field \\
HD168607d   & 5.19 & 8.37 &  \textcolor{blue}{18455}  & 2.81 &  \textcolor{blue}{10.1}  &  \textcolor{blue}{100}  &  \textcolor{blue}{45}, \textcolor{blue}{45}, \textcolor{blue}{35}  &  \textcolor{blue}{100}  & 12      &  \textcolor{blue}{45,36,36}  &    &  field  \\
MWC930b   & 2.88 & 4.65 &  \textcolor{blue}{16293}  & 1.77 & 7.2 &  \textcolor{blue}{100} &  2, 2, 2  & \textcolor{blue}{79}  & 18      &  \textcolor{blue}{14,13,13}  &     &  field \\
MWC930c    & 1.31 & 2.11 & 7411 & 1.77 &  \textcolor{blue}{$>$8}  &  \textcolor{blue}{46} &  1, 1, 0  & \textcolor{blue}{27}  & 18      & 3,3,3 &   \textcolor{blue}{B}   &  field \\
MWC314b  &     &  $<$1 mas  & 1.22 & 2.06 &    &    &  & $<$1     & 16      & $<$1,$<$1,$<$1  &  F   &  bound\\
MWC314c  & 1.18 & 1.9 & 5722 & 2.06 &  3.3 - 4.2  &  \textcolor{blue}{29} &  2, 1, 1  & 8 & 16      & 4,4,4 &      &  bound \\
HD152234b   & 0.51 & 0.83 & 1590 & 3.24 &    &   \textcolor{blue}{14} &  7, 7, 0  &  \textcolor{blue}{12}  & 10 & 1,1,1 &   G  &  borderline \\
WR102e$\_{b}$    & 0.71 & 1.15 & 9181 & 0.77 &  2.6 - 2.6  & 10 &   \textcolor{blue}{100},  \textcolor{blue}{94},  \textcolor{blue}{51}  & \textcolor{blue}{43}  & 42 & 4,4,4 &  \textcolor{blue}{B}  &  field  \\
HD164794b   &    &  2-8 mas  & 14 & 4.48 &    &    &  & $<$1     & 8 & $<$1,$<$1,$<$1  &  H   &  bound \\
HD164794c   & 2.4 & 3.87 & 6129 & 3.92 & 7.7 &   \textcolor{blue}{86} &   \textcolor{blue}{100}, \textcolor{blue}{100},  \textcolor{blue}{100}  & \textcolor{blue}{57}  & 8 &  \textcolor{blue}{16,15,15}  &     &  field  \\
HD164794d   & 3.5 & 5.65 & 8938 & 3.92 &   \textcolor{blue}{8.5}  &   \textcolor{blue}{100} &   \textcolor{blue}{100},  \textcolor{blue}{100},  \textcolor{blue}{100}  & \textcolor{blue}{83}  & 8 &   \textcolor{blue}{35,29,30}  &    &  field \\
HD164794e   & 4.6 & 7.42 &  \textcolor{blue}{11748}  & 3.92 &   \textcolor{blue}{8.7}  &   \textcolor{blue}{100} &   \textcolor{blue}{100},  \textcolor{blue}{100},  \textcolor{blue}{96}  & \textcolor{blue}{96}  & 8 &  \textcolor{blue}{61,45,45}  &    &  field \\
\hline                             
 \end{tabular}
\label{tableprobacomp}
\begin{flushleft}
Column details:\\
col. 1: star ID, col. 2: projected separation as measured in the image, cols. 3 and 4: probable deprojected separation in arcsec or in au, col. 5: maximum projected separation in arcsec corresponding to 10,000 au, col. 6: magnitude difference, col. 7$^{1}$: chance projection probability P(field) in NACO images, col. 8: P(field) using the trilegal model results and with counting from no restriction to $\pm$1$^{m}$ difference with the main star, 
col. 9: chance projection probability P$^{\prime}$(field) following the formula of \cite{correia2006} with trilegal data, col. 10: cut-off probability limit for P$^{\prime}$ (see text), col. 11$^{2}$: chance projection probability comparison P(field),  P$^{\prime}$(field), with P$^{\prime\prime}$(field) following the formula of \cite{ciar1999} and using 2MASS data. 
Col. 12: Proper motion comment, col. 13: status of the potential companion: ``bound'', ``borderline'', ``field''. 
\\
Table notes:\\
1: Probable separation should be $\le$10,000 au.\\
2: P(field) $\le$10\%.\\
3: Probabilities should be $\le$5\%.\\
4: ``real'' stands for companion in the disk plane with an inclination angle of 60$\degr$ \citep{ohara2003,smith2007} corresponding to a deprojection of 30$\degr$.\\
*: the separation goes up to 5205 au if the largest distance is considered. **: the separation goes up to 3725 au.\\
A: no motion estimates,
B: large motion indicating a foreground object,
C: small motion,
D: only SINFONI ``image'',
E: small motion, low projected speed,
F: see \cite{lobel2013}, G: only VISIR image, H: see \cite{rauw2012}.\\
The error in the chance projection probability estimates is usually lower than 1\% (see text).
\end{flushleft}
}
\end{table*}

\subsection{Method of chance projection probability}
\label{proba}
With a single epoch image, one can obtain an indication of whether a companion is bound to the main star by using the chance projection probability as previously performed by \cite{oud2010} with NACO for B and Be stars. 

As inputs for this determination, it is necessary to know
\begin{enumerate}
\item the projected distance of the potential companion to the main star in arcsec. It gives the projected area \textit{s}.
The probability of having an apparent stellar association is related to the possibility
that a star randomly falling in the area s can obtain the same image of the main star
with a companion;

\item the projected size of the field of view that provides the total projected area \textit{S} of the field of view. The ratio between \textit{S} and \textit{s} gives the number of area components \textit{N} of the system;

\item the magnitudes of the stars in the field of view or their flux ratio to the main star;

\item the Galactic coordinates of the main star;

\item the extinction coefficient A$_{v}$ of the main star;

\item the number of stars \textit{n} in the field of view. 
\end{enumerate}

There are two ways to determine \textit{n}.
The first  is a simple counting in the NACO (or another instrument) images.
The second  is to use a model of the Galactic star population, here we used the ``trilegal'' model \citep{trilegal}.
With the trilegal code, it is necessary to know point number 3 to provide a cut-off in magnitude, point 4 for the location in the Galaxy, point 5 to provide the corresponding extinction, and point 2 for the size of the field to consider.
The magnitude cut-off (20-21) in the K or Lp band is provided by the faintest star detected in NACO (aperture magnitude obtained with Sextractor).
We used different A$_{v}$ values in order to determine a range of solutions.

When K and Lp NACO images are available the results with the trilegal model with the K or Lp cut-off were compared and found to be identical. 
However, we also found that this model is not perfectly able to reproduce the star population in a few cases by comparing with the counting in the NACO images.
For instance in the field of the Pistol Star, the trilegal model appears to be too optimistic with excessive numbers of stars. Changing the value of A$_{v}$ does not help to diminish the number of stars. In the case of MWC314 and MWC930, the results appear too pessimistic with insufficient stars with respect to the NACO images. Indeed, changing the values of A$_{v}$ or the magnitude cut-off did not improve our comparison.

We consider the possibility of randomly finding a star in the area determined in point 1.
This defines the probability of chance projection that the potential companion is a field object and is not bound to the main star.
This probability P(event=k) with the previously defined parameters uses the combinatorics and the hypergeometric distribution \citep{stats} of the available stars once projected in the field

\begin{equation}
P(e=k)=\frac{C1 \times C2}{C3}
\label{eq1}
\end{equation}
with C1=C$^{k}_{m}$, C2=C$^{n-k}_{N-m}$, and C3=C$^{n}_{N}$.
The variables are   \textit{m}= number of companions visible in the image, \textit{k}=number of events for being a binary (corresponding to the number of stars to be added in the multiple system),
\textit{N}= number of area components, and \textit{n}= number of objects to distribute in \textit{N}.
The combinatorics function C corresponds to
\begin{equation}
C^{b}_{a}=(^{a}_{b})=\frac{a!}{b!(a-b)!}
\label{eq2}
\end{equation}





The chance projection probability with the different input sets is provided in Table~\ref{tableprobacomp} in columns 7 and 8. An estimate of the errors was performed considering a 5\% uncertainty on the counting of the number of stars in the field and/or in the area. It usually results in an uncertainty of less than 1\% in the chance projection probability. However, this error follows the separation as it increases, but this does not affect the interpretation of the probability results.

We list the projected separation of the main star to the potential companion in arcsec. 
The NACO image is a projection and the potential companion stars can be above or below the plane of projection. Thus, we account for the deprojected distance by correcting the projected distance with the most probable angle for the random distribution of the orbital plane inclination angle ($\arcsin(\pi/4)$).
In the case of HD168625 the inclination angle is known and we present both estimates using the known angle (real) and the most probable angle.
We adopt a significance level of  10\%  for the probability of random association to consider that a companion is physically bound to its central star.
This threshold of 10\% corresponds to a  slightly significant result (1.7 $\sigma$), while 5\% would be considered more significant (2 $\sigma$).

In  addition to the formula we used, the probability P$^{\prime}$(field) that a star is a field object can be computed following \cite{correia2006}.
 For PMS objects they consider that a probability of 1\% is not negligible and the nearby star should be considered as a field star (the association only results by a chance alignment).
However, the distance to Earth of the objects is not taken into account by their formula. The objects in their article are at a distance below 190 pc. Therefore, by scaling this formula, and the cut-off for being bound or not to the distance of the star to Earth, we compute a cut-off probability ranging from 8\% at 1,580 pc to 62\% at 11,800 pc.
In Table~\ref{tableprobacomp} in columns 9 and 10, we also report  the probability of a chance projection following the method of \cite{correia2006} and the scaled cut-off probability. We also note that our formula is usually more pessimistic (70\% of the time), i.e. providing a higher probability value than the formula by \cite{correia2006} using the same type of counting with the trilegal data.

To obtain another estimate, we used the probability P$^{\prime\prime}$(field) defined by \cite{ciar1999}, which is given  in col. 11 in Table~\ref{tableprobacomp}. 
We compare the results of this method with the probabilities obtained with our formula and with the method of \cite{correia2006} using the same 2MASS raw data. The result is also provided in col. 11 following  the order: our formula, the  \cite{correia2006} formula, and the  \cite{ciar1999} formula.
The results are  comparable even though, as already noted, their probability of chance projection is often a bit lower than the one determined with our method.
\cite{ciar1999} consider two objects bound if the probability of a chance superposition is less than 5\%.
However, as illustrated with the Pistol Star in Fig.~\ref{FigPS} for very far objects, non-AO (2MASS-K) images result in a lower number of visible stars than with AO-assisted images (NACO-K). As a consequence, chance projection probability determination should be performed with high angular resolution images or adequate catalogues for those distant stars.


\subsection{Multi-epoch observations and proper motions}
\label{MEO}

Multi-epoch images can be used to determine potential proper motions of the objects and eventually an orbit, or to detect whether the star is a background or foreground star. If the motion of the ``neighbour'' star is larger than the motion of the main object then it is considered to be a foreground field object. If the motion is of the same order of value or smaller than the main star, then the neighbour is considered to be at the same distance of the main star or to be a background object. However, the proper motion of the main star is usually not known or is uncertain. The difference in right ascension and declination, but also in separation between the main star and the potential companion, is considered in order to  detect any proper motion. 
We mainly considered the separation difference because in most of the cases the sampling or the signal-to-noise ratio is not high enough to safely estimate a position to within a fraction of pixel.
We only consider valid a proper motion larger than the size of the biggest pixel in the sample for each star.
The absolute-coordinate comparison was discarded to avoid systematics and shifts between the  World Coordinate System (WCS) calibration of images originating from different cameras.
Considering the large distance to the stars, providing an orbit goes beyond the possibility of the data.

To perform a multi-epoch analysis, we require good spatial resolution and a pixel size that is as small as possible.
For instance, considering a displacement of 10 mas at the speed of 10 \kms~at 2 kpc, both epochs should be separated by an interval of 10 yrs.
With a displacement of 50 mas at the speed of 100 \kms~at 8 kpc, the two epochs should be separated by an interval of 19 yrs.
Both examples consider the best possible case with a motion parallel to the projection plane.

Using the Kepler and Newton laws for different mass ranges of a circular orbit with different star separations (up to 10,000 au), we obtain a range of possible orbital velocities and periods. From the simulations, the orbital velocities should be less than 10 \kms~with periods of thousands or tens of thousands of years. It is believed that higher velocities indicate field objects.  

To obtain larger time intervals, we searched in  the ESO and MAST archives.
The best archival images in terms of spatial resolution and pixel size are those from the HST-NICMOS and WFC for the objects of our sample.
HST images were retrieved for the Pistol Star, WR102e, HD168625, HD168607, MWC930, LBV1806-20, and HD164794.
In the last case, the images are not used because they are underexposed for our purpose.
The proper motion of objects found in the field of LBV1806-20 and HD168607 are not discussed here because they are too distant ($>$10,000 au) or too faint ($\delta$mag=10$^{m}$) to  be bound.
Unfortunately, there is no HST data for MWC314 and the EMMI images do not show the nearby star at 1.18$\arcsec$.
For the stars HD168625, the Pistol Star, WR102e, and MWC930, the characteristics of their multi-epoch observations are provided in Table~\ref{sumepochs}. Table~\ref{sumPM} compiles the proper motion information and the implied binary status for stars HD168625, the Pistol Star, WR102e, and MWC930.

\begin{table}[h]
\tiny{
\centering
\caption[]{Summary of the characteristics of the multi-epoch observations for several stars in our sample.
The epochs in MJD and the civil date are provided at the bottom of the table.}
\centering
\begin{tabular}{@{\ }l@{\ \ }l@{\ \ }l@{\ \ }l@{\ \ }l@{\ \ }l@{\ \ }}
\hline 
Star    &       Epoch   &       Instrument      &       Time int. &     Pixel size &  Limiting        \\
        &               &               &        &       &      displacement,   \\
        &               &               &       yr &    mas &   velocity        \\
\hline                                                                                  
Pistol Star     &       1a$^{*}$-2a     &       NICMOS-NACO     &       12.76   &       13-50   &       50 mas     \\
and     &       2a-3a   &       NACO-NACO       &       1.08    &               &       150 \kms    \\
WR102e  &       1a-3a   &       NICMOS-NACO     &       13.84   &               &               \\
HD168625        &       1b-2b   &       HST/WF-WF       &       0.19    &       27-50   &       50 mas     \\
        &       2b-3b   &       WF-NICMOS       &       1.07    &               &       40 \kms    \\
        &       3b-4b   &       NICMOS-NACO     &       12.88   &               &        at 2.2 kpc      \\
        &       1b-4b   &       WF-NACO &       14.14   &               &               \\
MWC930  &       1c-2c   &       HST/ACS$^{**}$–NACO   &       5.15    &       25-27   &       27 mas, 80 \kms    \\
\hline                                                                                  
\end{tabular}
\label{sumepochs}
\begin{flushleft}
*: Epoch 1a from \cite{figer1998}, **: \cite{acs}\\
Epochs ``a'': 1a=50705.41-14/09/1997, 2a=55364.23-17/06/2010, 3a=55760.05-18/07/2011. \\
Epochs ``b'': 1b=50597.59-29/05/1997, 2b=50666.80-06/08/1997, 3b=51056.06-31/08/1998, 4b=55761.17-19/07/2011. \\
Epochs ``c'': 1c=53880.35-25/05/2006, 2c=55761.06-19/07/2011.
\end{flushleft}
}
\end{table}

\begin{table}[h]
\tiny{
\centering
\caption[]{Summary of the proper motion estimates and inferences about possible membership in a binary system of potential companion as defined in Table~\ref{tableprobacomp}.}
\centering
\begin{tabular}{@{\ }l@{\ \ }l@{\ \ }l@{\ \ }l@{\ \ }l@{\ \ }}
\hline 
Star    &       Neighbour       &       Displacement    &       Comments        &       Conclusion      \\
&               &       mas     &               &               \\
\hline                                                                  
Pistol Star     &       P4, P2  &               &       not enough data &               \\
        &       P3      &       250     &       too large       &       foreground      \\
        &       P1      &       27      &       smaller than pixel      &               \\
        &       P8,P9,P10,P13   &       $>$ 50  &       too large       &       foreground      \\
        &       P14,P15,P16     &       $>$ 50  &       too large       &       foreground      \\
WR102e  &       b       &       $>$ 50  &       too large       &       foreground      \\
HD168625        &       b       &       $<$ 27  &       approaching a   &       bound   \\
        &       d, k, m &       51 to 161       &       too large       &       foreground      \\
        &       j, h    &       $<$50   &       small   &       background      \\
        &       c       &       50      &       away from a     &       not bound   \\
MWC930  &       c       &       $>$ 80  &       too large       &       foreground      \\
\hline                                                                  
\end{tabular}
\label{sumPM}
}
\end{table}

The motion found for HD168625b at 1.15$\arcsec$ (see Fig.~\ref{Fighd168625b}) is smaller than the NACO pixel size. If true, it gives a projected velocity in the image of less than 10 \kms. Its motion could indicate that it is  approaching HD168625a. This is one of the rare cases where we can actually obtain an estimate of a fraction of a pixel for the position of the star. However, this was not possible in some HST images.
Recently, \cite{aldo2014} confirmed by interferometric observations the presence of this companion.

Star c has a motion of the order of 50 mas going away from HD168625. 
The other stars are only visible in our ``wide'' NACO image.


\subsection{Star separation and flux ratio tests}
\label{sepa}
Because the separation is projected in the images, we de-project to derive the proper distance of the potential companion to the main star. For this purpose the deprojected distance is obtained using the most probable angle in the case of random distribution of the orbital inclination angle (corresponding to $\arcsin(\pi/4)$). If the distance of the main object to Earth is known,  the separation in au of the potential companion to the main star  can be determined. For wide binaries it can range up to 20,000 au \citep{lepine2007,long2010} or even larger \citep{shaya2011}, but we use a conservative distance of 10,000 au. It corresponds to about 90\% of the binary star distribution shown in \cite{lepine2007}. \cite{krum2009} and \cite{krum2012} also indicate that companions can be expected at a distance of 5,000 au or more.
\cite{sana2014} show that at a separation of 8$\arcsec$, the cumulative multiplicity  frequency reaches 91\%. It would correspond to a distance of 16,000 au at a typical distance of 2 kpc.
For deprojected distances above 10,000 au the neighbour star is probably not bound to the main star \citep[][actually indicate that at large separation only the brightest companions are bound to the main star]{sana2014}. The maximum projected separation in arcsec corresponding to the linear separation of 10,000 au in the image is also provided (see Table~\ref{tableprobacomp}). The results are indicated in Table~\ref{tableprobacomp}, cols 3 to 5.
Stars P4; HD168625b; HD168607b; MWC314b,c; MWC930c; WR102e$_{b}$; HD152234b; and HD164794b,c,d are separated from the main component by less than 10,000 au. This test eliminates  50\% of the potential companions visible in the images as bound  in a binary system. 

According to the study by \cite{sana2014} among massive stars, it is not rare to find binaries with a magnitude difference in K or Lp of $\sim$8-8.5$^{m}$. The limiting magnitude would ensure  that almost all potential companions are caught. 
In addition, they also used NACO  to find binaries in O stars, as  this threshold would be adequate for LBV stars and corresponds to a late B-type star \citep[using the models from ][]{schaller1992}.
The results are provided in Table~\ref{tableprobacomp}, col. 6, while we give  the angular projected separation corresponding to 10,000 au as an indication in col. 5. This test allows us to discard 30\% of the potential companions visible in the images as bound in a binary system.
The results are discussed in Section~\ref{bin}.


\subsection{Binarity assessment and discussion}
\label{bin}

To summarize, a companion is bound to the main star if it complies with the following criteria:
\begin{itemize}
\item The deprojected/probable separation must be smaller than 10,000 au.
This value is given in Table~\ref{tableprobacomp}, col. 4. The projected separation (from col. 2) is corrected with the most probable angle of projection to provide a more realistic angular separation (col. 3) converted to au in column 4.
\item The difference in magnitudes must be $\delta$mag$<$8.5$^{m}$ (Table~\ref{tableprobacomp}, col. 6).
\item The chance projection probability P(field) must be smaller than 10\%,  P$^{\prime}$(field) smaller than the cut-off probability, or P$^{\prime\prime}$(field) smaller than 5\%. 
In column 7 we indicate the chance projection probability for the potential companion considering the stars counted in the NACO images (except for WR102ka for which SINFONI image is used). 
In column 8, we provide the chance projection probability for the potential companion considering the stars counted in the results of simulations with the trilegal code \citep{trilegal}. They differ by the magnitude ranges used: all trilegal stars, stars with a maximum difference of $\pm$2$^{m}$ and $\pm$1$^{m}$. 

In column 9, the chance projection probability following the formula of \cite{correia2006} is provided and the scaled cut-off probability is given in column 10.
In column 11, the chance projection probabilities P(field), P$^{\prime}$(field), and P$^{\prime\prime}$(field)  respectively following our formula, the formula of \cite{correia2006}, and the formula of \cite{ciar1999} are provided using the same set of 2MASS data. The same 5\% limit is applied to the result of  P(field) and P$^{\prime}$(field) in col. 11 for consistency  with the result of P$^{\prime\prime}$(field). When they show contradictory results, we consider the majority of results as a better trend of the probability. 
\item The multi-epoch analysis must yield a small motion (Table~\ref{tableprobacomp}, col. 12). 
\end{itemize}

The conclusion about the status of the potential companion, bound,   not bound,  or borderline is given in the last column of Table~\ref{tableprobacomp}.
\begin{itemize}
\item Therefore, HD168625 and MWC314 have a companion in a wide orbit. The latter is known to host a close companion at less than 1 mas \citep{lobel2013} that is beyond the capability of the imaging technique. MWC314 could be a triple hierarchical system like HD152234 \citep{sana2008}.
HD168625b being bound to HD168625a would imply that it is in the equatorial plane of the system inclined by 60$\degr$ with respect to the observer. The role of this star is questionable for shaping the inner shell, which appears somewhat distorted.
Could this companion help to clear the material ejected by the central star in the inner cavity?
\item P4 is a borderline companion for the Pistol Star, similarly for HD152234b with HD152234a.
HD152234b is probably bound to the main star because the separation is small, the magnitude ratio is 2.5 according to \cite{mason2009}, and the chance probability is low but at the limit.  Based on the observations of \cite{mason2009}, \cite{sana2008} also concluded that this companion is probably bound to the main star. Together with the central spectroscopic binary it composes a triple system.
\end{itemize}

In addition, some stars are known to host a close companion not resolved by the imaging techniques. This is the case for the Pistol Star \citep{martayan2012}; MWC314, see above; and LBV1806-20, whose spectrum showed a double line signaling a close companion \citep{figer2004} estimated by \cite{eiken2004} to orbit at less than 450 au (representing much less than 1 mas). However, there is currently no spectroscopic monitoring to confirm the binary nature.
In addition, \cite{rauw2012} found a close companion at $\sim$10 mas of HD164794, while \cite{sana2014} resolved the system with the VLTI-PIONIER \citep{pionier} and found a separation of 5 mas. \cite{aldo2014} estimated the separation to be $\sim$19 mas. 
These companions are not ruled out by our criteria.
Overall, about two out of seven of the LBV stars in our sample are found to host a companion in a wide orbit. 


Because more than 75\% of O stars are binaries according to \cite{chini2012} and \cite{sana2012,sana2014}, it is expected that LBVs are binaries as well, but very few LBVs are known to host a companion \citep{rivinius2014,lobel2013,dam1997}. Following \cite{sana2014}, about 30-35\% of the binaries are found with a large projected separation (as in this paper). 
In this study, two (HD168625, MWC314) out of the seven LBVs (29\%) in the sample with high angular resolution images are binaries with a companion in a wide orbit. Because of the huge luminosity of the LBVs, it is quite difficult to detect the surrounding companions. It implies that the mutliplicity rate found is certainly a lower limit.

The Pistol Star, WR102ka, MWC930, HD168607, and LBV1806-20 do not seem to have a companion on a wide orbit, even if P4 is borderline for the Pistol Star. However, the statistical analysis was performed on a small number of stars and should be extended to more LBV/cLBVs, although the sample we use already represents 20\% of the stars of this type \citep[considering the sample of 35 LBVs + cLBVs defined by ][]{clark2005}. 
Recently, \cite{aldo2014} resolved or partially resolved  a companion around three out of nine Galactic LBVs, producing a binary rate of 33\% in their sample. 
Among the LBVs, only HD168625 is shared by both studies. 
Joining the two studies, the sample results in 4 binaries among 15 LBVs. It corresponds to a 27$\pm$17\% binary fraction with a confidence rate of 95\%.

What could be the impact of a companion in a wide orbit on the evolution of the central star and its eruptions?
With a separation of at least 1500 au, unless with a highly eccentric orbit, they appear too  distant to trigger any eruption by mass transfer or to have any impact on the day-to-day stellar evolution of the main star. 
In such a case, their main role would be to allow the formation of the  star and the nebula \citep[see][]{krum2012} in addition to the possible shaping of the nebula.
However, according to \cite{kaib2014}, despite a large separation of thousands of au, collisions could occur between the components of a binary system on average every 1,000 to 7,500 yrs in our Galaxy. If this is true it could imply that despite a large separation, companions in a wide orbit could still trigger cataclysmic events or eruptions. 

\section{General circumstellar environment description}
\label{GCED}

\begin{table}[h]
\centering
\caption[]{Detected nebular shells, their semi-minor and semi-major dimensions  in arcsec and in pc. In the case of a round shell, only one extent is provided. 
The extent size is defined using the convention of  10\% of the peak surface brightness  \citep{PN2003}.
}
\centering
\tiny{
\begin{tabular}{@{\ }l@{\ \ }l@{\ \ }l@{\ \ }l@{\ \ }l@{\ \ }l@{\ \ }l@{\ \ }}
\hline 
Star & shell radii  & shell radii & $\lambda$ & Figure & shell radii \\
 & arcsec $\times$ arcsec & pc $\times$ pc & $\mu$m & & pc $\times$ pc \\
 &  &  &  & &  \cite{weis2011} \\
\hline
Pistol Star &   25.5 $\times$ 24            & 0.99 $\times$ 0.93 & 8 & \ref{FigPS} & 0.8 $\times$ 1.2\\
WR102ka &         58 $\times$ 55          & 2.25 $\times$ 2.13 & 22 & \ref{Figwr102ka} &\\
LBV1806-20* &      \_         & \_ & 3.4-22 & \ref{Figlbv1806} &\\
HD168625 int$^{1}$&     19 $\times$ 14           & 0.20 $\times$ 0.15  & 22  & \ref{Fighd168625b}, \ref{Fighd168625IR} & 0.17 $\times$ 0.13\\
HD168625 imd$^{1}$&      46 $\times$ 41          & 0.49 $\times$ 0.44  & 22 & \ref{Fighd168625IR} &\\
HD168625 ext$^{1}$&      $ 80: \times 70: $          & 0.85 $\times$ 0.75  & 22 & \ref{Fighd168625IR} &\\
HD168625 ring$^{1}$&      $33 \times 40$ & 0.35 $\times$ 0.43  & 8 & \ref{Fighd168625IR} &\\
HD168607 &        9.5           & 0.10 & 22  & \ref{Fighd168607} &\\
MWC930 &         59 $\times$ 54          & 1.00 $\times$ 0.92 & 22 & \ref{Figmwc930} & \\
MWC314   &        26           & 0.38 & 22 & \ref{Figmwc314} &\\
HD152234 &         16-20           & 0.15-0.19 & 3.4-22 & \ref{Fighd152234} &\\
HD164794 &        11          & 0.06-0.10 & 5.6  & \ref{Fighd164794} &\\
\hline
\end{tabular}
\label{sumsize}
\begin{flushleft}
The distance of the stars to the Earth is available in Table~\ref{starsample}. \\
*: There is no indication for the star LBV1806-20 because it is deeply embedded in a complex with other WR stars and it is not possible to determine to which star the nebula belongs.\\
The last column provides the measurements performed by \cite{weis2011} in pc.\\
1: The outer radius measured along the axes ESE-WNW and NE-SW of the internal (int), the intermediate (imd), and the external (ext) shells are given for HD168625. As a reference, the dimensions of the NE ring are also indicated.
\end{flushleft}
}
\end{table}

In the infrared images, all stars in the sample exhibit surrounding nebular shells. Their size was measured in the images using the convention of  10\% of the peak surface brightness as defined in planetary nebula studies \citep{PN2003}. Taking into account their distance to Earth, the shell size in parsec (pc) is listed in Table~\ref{sumsize}.  
The Pistol
Star, HD168625, MWC930, and WR102ka all exhibit large shells with a diameter of at least 1 pc. MWC314 presents an intermediate-size shell of about 0.8 pc diameter, but we cannot detect in the infrared the large bipolar structure found by \cite{marston2008} in H$\alpha$. 
For reference, $\eta$ Car's shell ``radius'' is about 0.5 pc.
We do not provide any measurement for LBV1806-20 because it is deeply embedded in a complex nebular environment with several WR stars. It is therefore not possible to safely assign shell membership to a given star.
In some cases, the envelope is smaller than 0.2 pc and might result from another mechanism than matter ejected by giant LBV eruption. This is the case for the non LBVs HD152234, HD164794, and possibly for the LBV HD168607.
In Section~\ref{hd168625shells} the environment of HD168625 and its multiple shells is discussed as an example, while in Appendix~\ref{GCED2} the main environmental structures of the remaining stars of the sample are presented along with their corresponding infrared images.

\subsection{HD168625: shells and rings}
\label{hd168625shells}


    \begin{figure*}[h!]
   \centering
   \includegraphics[width=\hsize]{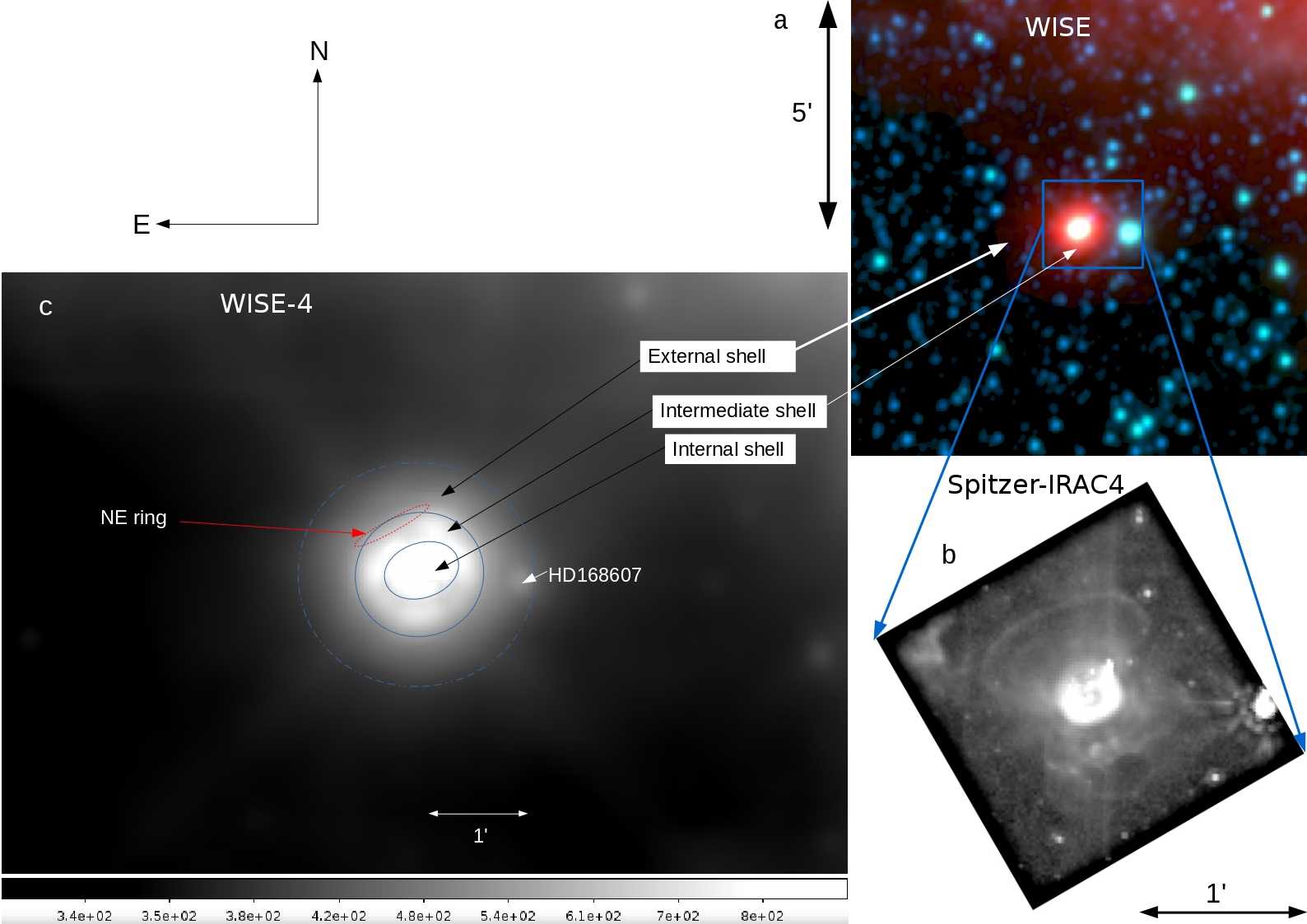}
      \caption{Star HD168625, Spitzer and WISE images:
      a) WISE $10\arcmin \times 10\arcmin$ tricolour image R=WISE4, G=WISE2, B=WISE1. The intermediate and external nebulae are visible. The star $\simeq 1\arcmin$ west of HD168625 is HD168607.
      b) $2.75\arcmin \times 2.75\arcmin$ Spitzer-irac4 image. The image shows the nebular rings reported by \cite{smith2007}. 
      c) WISE channel 4 image at 22 $\mu$m of HD168625 with labels of the internal, intermediate, and external shells (shown by ellipses). The position of the NE ring reported by \cite{smith2007} is also indicated. It matches the outer border of the intermediate shell and the inner border of the external shell. In these images, it is just possible to see a local flux minimum between the intermediate and external shell. This flux minimum is used to define the limits of both shells.
      }
         \label{Fighd168625IR}
   \end{figure*}

In the infrared images (Figs.~\ref{Fighd168625b}, \ref{Fighd168625IR}), HD168625 clearly shows two different shells and possibly an additional fainter third component. Their sizes (in arcsec and in pc) are listed in Table~\ref{sumsize}.
The internal shell is well known and we observed it with NACO (Fig.~\ref{Fighd168625b}). Depending on the wavelength,  different extents of the same shell can be observed (i.e. NACO versus WISE).
The dynamical age of the internal shell ranges from 4,000 to 5,700 yr according to \cite{nota1996}, \cite{pasquali2002}, and \cite{ohara2003}.

By doing a cut of the WISE4 image or using the flux gradient,  a break in the latter can be seen with two different slopes.
The steeper one corresponds to the known ``intermediate'' shell \citep[][their ``outer shell'']{hut1994}, while the other one in the outer part could correspond to a fainter redder external shell. Its inner part would be blended with the intermediate shell (see Fig.~\ref{Fighd168625IR}). 
A forthcoming study will present  this external shell in detail; in particular, careful comparison with the WISE4 extended PSF\footnote{http://wise2.ipac.caltech.edu/docs/release/allsky/expsup/\\sec4\_4c.html\#psf} and features\footnote{http://wise2.ipac.caltech.edu/docs/release/allsky/expsup/sec2\_4bii.html\\
\#conspicuous\_PSF\_ring} was performed to confirm the presence of this external shell. 
Interestingly, the famous rings discovered by  \cite{smith2007} in the IRAC4 image coincide with the inner border of the external shell (see Fig.~\ref{Fighd168625IR}). 
He indicates that the internal equatorial ring and the intermediate shell are coeval with an age of a few thousand years.
The intermediate and external shells probably result from previous eruptions or stellar phases with wind expansion.

Ghost patterns in the MIPS image prevent us from  properly finding the extent of the external shell.
The ratio of the semi-major to the semi-minor axis ranges from 1.1 to 1.2 for the external and intermediate shells, respectively, while it ranges from 1.4 to 1.5 for the internal shell.

If the external shell is confirmed, it could indicate that the rings of HD168625 are possibly formed by the collision of a fast wind with a slower and previously ejected material. This observation supports the scenario proposed by \cite{chita2008}. According to these authors, the formation of such rings could be the result of the collision of an anisotropic wind created during the blue supergiant phase of a fast rotating star with a shell created by the star in a previous red supergiant phase. 
\cite{smith2007b} proposed  that the rings are formed in the blue supergiant phase. 
\cite{smith2013} suggested that the collision with the fast wind of a blue supergiant is not needed to form the rings. An expanding equatorial ring may be ejected by a blue fast rotating supergiant star with a clearing of the caps. This scenario does not require the presence of another older shell.

Other objects like SN1987A or Sher 25 also reveal similar rings.
\cite{hendry2008} ruled out the presence of a binary companion and found that the rings of the blue supergiant Sher 25 were formed during the blue loop evolutionary phase. 
The rings of SN1987A have shown brightening episodes explained by the collision of matter with different velocities \citep{larsson2011}.

Another alternative scenario is proposed by \cite{morris2009} who invoked the merger model to explain the creation of the rings during the red supergiant phase. However, it only explains co-eval shells.

The exact nature of stellar fast and slow outflow is unknown. Do they originate from a normal stellar wind of a red or blue supergiant or from LBV eruption?
HD168625 could become a template star that explains the other stars with similar rings such as Sher 25.



\section{Conclusions}
\label{conclue}
This study, based on multiwavelength imaging techniques, allowed  
the environment of LBVs and other massive stars to be probed on a scale of $\sim$0.1$\arcsec$.
Our conclusions are the following:

   \begin{enumerate}
      \item In our sample, two out of seven of the LBVs could have a companion in a wide orbit.
      \item Several stars could be a triple hierarchical system with a close and a wide binary system (possibly MWC314 or HD152234).
      \item For the first time LBV companions were directly imaged.
      \item The presence of a companion in a wide orbit is an indicator of the way the star formed: as a group of stars. It also contributes to constraining models of massive star formation. However, the actual contribution of the wide orbit companion to the main day-to-day stellar evolution of the central star is questionable.
      \item All stars are surrounded by circumstellar matter, ejected in outbursts,  by a wind, or both.       
      \item The creation of the nebular rings of HD168625 is possibly due to the collision of fast material with a previous slower shell, which is presented here for the first time but needs to be confirmed. 
      \item The companions can act on the shaping of the surrounding nebula. HD168625b could be at the origin of the clearing of the internal cavity and of the asymmetric structure of the internal ring.
      \end{enumerate}
Other spatial scales must be probed with other techniques (interferometry and spectroscopy) to infer the presence of close companions and to analyse the circumstellar matter structure close to the star. 


\begin{acknowledgements}
We thank the anonymous referee and the editor Dr R. Kotak for their helpful comments that allowed us to improve the manuscript.
CMa is grateful to ESO for the allocation of 2 months of a temporary reassignment to the science office.
A.L. acknowledges financial support of the ESA-Prodex Programme administered by the Belgian Science Policy Office.
CMa thanks J. Taylor for providing the Reflex prototype of the VISIR dataflow.
This research has made use of the Simbad and VizieR databases maintained at the CDS, Strasbourg, France, of NASA's Astrophysics Data System Bibliographic Services, and of the NASA/IPAC Infrared Science Archive, which is operated by the Jet Propulsion Laboratory, California Institute of Technology, under contract with the US National Aeronautics and Space Administration. 
This publication makes use of data products from the Two Micron All Sky Survey, which is a joint project of the University of Massachusetts and the Infrared Processing and Analysis Center/California Institute of Technology, funded by the National Aeronautics and Space Administration and the National Science Foundation.
This research has made use of the NASA/ IPAC Infrared Science Archive, which is operated by the Jet Propulsion Laboratory, California Institute of Technology, under contract with the National Aeronautics and Space Administration. Some of the data presented in this paper were obtained from the Mikulski Archive for Space Telescopes (MAST). STScI is operated by the Association of Universities for Research in Astronomy, Inc., under NASA contract NAS5-26555. Support for MAST for non-HST data is provided by the NASA Office of Space Science via grant NNX13AC07G and by other grants and contracts. Based on observations made with the NASA/ESA Hubble Space Telescope, and obtained from the Hubble Legacy Archive, which is a collaboration between the Space Telescope Science Institute (STScI/NASA), the Space Telescope European Coordinating Facility (ST-ECF/ESA) and the Canadian Astronomy Data Centre (CADC/NRC/CSA).
This publication makes use of data products from the Wide-field Infrared Survey Explorer, which is a joint project of the University of California, Los Angeles, and the Jet Propulsion Laboratory/California Institute of Technology, funded by the National Aeronautics and Space Administration. This work is based in part on observations made with the Spitzer Space Telescope, which is operated by the Jet Propulsion Laboratory, California Institute of Technology under a contract with NASA. 
This research has made use of the ESO archive.
Herschel is an ESA space observatory with science instruments provided by European-led Principal Investigator consortia and with important participation from NASA. This research made use of data products from the Midcourse Space Experiment. Processing of the data was funded by the Ballistic Missile Defense Organization with additional support from NASA Office of Space Science. 
\end{acknowledgements}



\bibliographystyle{aa}
\bibliography{article12bib}

\begin{thebibliography}{98}
\expandafter\ifx\csname natexlab\endcsname\relax\def\natexlab#1{#1}\fi

\bibitem[{{Aldoretta} {et~al.}(2015){Aldoretta}, {Caballero-Nieves}, {Gies},
  {Nelan}, {Wallace}, {Hartkopf}, {Henry}, {Jao}, {Ma{\'{\i}}z Apell{\'a}niz},
  {Mason}, {Moffat}, {Norris}, {Richardson}, \& {Williams}}]{aldo2014}
{Aldoretta}, E.~J., {Caballero-Nieves}, S.~M., {Gies}, D.~R., {et~al.} 2015,
  \aj, 149, 26

\bibitem[{{Ankay} {et~al.}(2001){Ankay}, {Kaper}, {de Bruijne}, {Dewi},
  {Hoogerwerf}, \& {Savonije}}]{ankay2001}
{Ankay}, A., {Kaper}, L., {de Bruijne}, J.~H.~J., {et~al.} 2001, \aap, 370, 170

\bibitem[{{Baumgardt} \& {Klessen}(2011)}]{baum2011}
{Baumgardt}, H. \& {Klessen}, R.~S. 2011, \mnras, 413, 1810

\bibitem[{{Blomme} \& {Volpi}(2014)}]{blomme2014}
{Blomme}, R. \& {Volpi}, D. 2014, \aap, 561, A18

\bibitem[{{Boggess} {et~al.}(1978){Boggess}, {Carr}, {Evans}, {Fischel},
  {Freeman}, {Fuechsel}, {Klinglesmith}, {Krueger}, {Longanecker}, \&
  {Moore}}]{iue1978}
{Boggess}, A., {Carr}, F.~A., {Evans}, D.~C., {et~al.} 1978, \nat, 275, 372

\bibitem[{{Bondar}(2012)}]{bondar2012}
{Bondar}, A. 2012, \mnras, 423, 725

\bibitem[{{Bonnet} {et~al.}(2004){Bonnet}, {Abuter}, {Baker}, {Bornemann},
  {Brown}, {Castillo}, {Conzelmann}, {Damster}, {Davies}, {Delabre},
  {Donaldson}, {Dumas}, {Eisenhauer}, {Elswijk}, {Fedrigo}, {Finger},
  {Gemperlein}, {Genzel}, {Gilbert}, {Gillet}, {Goldbrunner}, {Horrobin}, {Ter
  Horst}, {Huber}, {Hubin}, {Iserlohe}, {Kaufer}, {Kissler-Patig}, {Kragt},
  {Kroes}, {Lehnert}, {Lieb}, {Liske}, {Lizon}, {Lutz}, {Modigliani}, {Monnet},
  {Nesvadba}, {Patig}, {Pragt}, {Reunanen}, {R{\"o}hrle}, {Rossi}, {Schmutzer},
  {Schoenmaker}, {Schreiber}, {Stroebele}, {Szeifert}, {Tacconi}, {Tecza},
  {Thatte}, {Tordo}, {van der Werf}, \& {Weisz}}]{sinfo2}
{Bonnet}, H., {Abuter}, R., {Baker}, A., {et~al.} 2004, The Messenger, 117, 17

\bibitem[{{Brandt}(2014)}]{stats}
{Brandt}, S. 2014, {Data analysis : statistical and computational methods for
  scientists and engineers, Springer International Publishing, ISBN:
  9783319037622}

\bibitem[{{Chentsov} \& {Gorda}(2004)}]{chentsov2004}
{Chentsov}, E.~L. \& {Gorda}, E.~S. 2004, Astronomy Letters, 30, 461

\bibitem[{{Chini} {et~al.}(2012){Chini}, {Hoffmeister}, {Nasseri}, {Stahl}, \&
  {Zinnecker}}]{chini2012}
{Chini}, R., {Hoffmeister}, V.~H., {Nasseri}, A., {Stahl}, O., \& {Zinnecker},
  H. 2012, \mnras, 424, 1925

\bibitem[{{Chita} {et~al.}(2008){Chita}, {Langer}, {van Marle},
  {Garc{\'{\i}}a-Segura}, \& {Heger}}]{chita2008}
{Chita}, S.~M., {Langer}, N., {van Marle}, A.~J., {Garc{\'{\i}}a-Segura}, G.,
  \& {Heger}, A. 2008, \aap, 488, L37

\bibitem[{{Ciardullo} {et~al.}(1999){Ciardullo}, {Bond}, {Sipior}, {Fullton},
  {Zhang}, \& {Schaefer}}]{ciar1999}
{Ciardullo}, R., {Bond}, H.~E., {Sipior}, M.~S., {et~al.} 1999, \aj, 118, 488

\bibitem[{{Clark} {et~al.}(2005){Clark}, {Larionov}, \& {Arkharov}}]{clark2005}
{Clark}, J.~S., {Larionov}, V.~M., \& {Arkharov}, A. 2005, \aap, 435, 239

\bibitem[{{Correia} {et~al.}(2006){Correia}, {Zinnecker}, {Ratzka}, \&
  {Sterzik}}]{correia2006}
{Correia}, S., {Zinnecker}, H., {Ratzka}, T., \& {Sterzik}, M.~F. 2006, \aap,
  459, 909

\bibitem[{{Damineli} {et~al.}(1997){Damineli}, {Conti}, \& {Lopes}}]{dam1997}
{Damineli}, A., {Conti}, P.~S., \& {Lopes}, D.~F. 1997, \na, 2, 107

\bibitem[{{Dekker} {et~al.}(1986){Dekker}, {Delabre}, \& {Dodorico}}]{emmi1986}
{Dekker}, H., {Delabre}, B., \& {Dodorico}, S. 1986, in Society of
  Photo-Optical Instrumentation Engineers (SPIE) Conference Series, Vol. 627,
  Instrumentation in astronomy VI, ed. D.~L. {Crawford}, 339--348

\bibitem[{{Eikenberry} {et~al.}(2004){Eikenberry}, {Matthews}, {LaVine},
  {Garske}, {Hu}, {Jackson}, {Patel}, {Barry}, {Colonno}, {Houck}, {Wilson},
  {Corbel}, \& {Smith}}]{eiken2004}
{Eikenberry}, S.~S., {Matthews}, K., {LaVine}, J.~L., {et~al.} 2004, \apj, 616,
  506

\bibitem[{{Eisenhauer} {et~al.}(2003){Eisenhauer}, {Abuter}, {Bickert},
  {Biancat-Marchet}, {Bonnet}, {Brynnel}, {Conzelmann}, {Delabre}, {Donaldson},
  {Farinato}, {Fedrigo}, {Genzel}, {Hubin}, {Iserlohe}, {Kasper},
  {Kissler-Patig}, {Monnet}, {Roehrle}, {Schreiber}, {Stroebele}, {Tecza},
  {Thatte}, \& {Weisz}}]{sinfo1}
{Eisenhauer}, F., {Abuter}, R., {Bickert}, K., {et~al.} 2003, in Society of
  Photo-Optical Instrumentation Engineers (SPIE) Conference Series, Vol. 4841,
  Instrument Design and Performance for Optical/Infrared Ground-based
  Telescopes, ed. M.~{Iye} \& A.~F.~M. {Moorwood}, 1548--1561

\bibitem[{{Fazio} {et~al.}(2004){Fazio}, {Hora}, {Allen}, {Ashby}, {Barmby},
  {Deutsch}, {Huang}, {Kleiner}, {Marengo}, {Megeath}, {Melnick}, {Pahre},
  {Patten}, {Polizotti}, {Smith}, {Taylor}, {Wang}, {Willner}, {Hoffmann},
  {Pipher}, {Forrest}, {McMurty}, {McCreight}, {McKelvey}, {McMurray}, {Koch},
  {Moseley}, {Arendt}, {Mentzell}, {Marx}, {Losch}, {Mayman}, {Eichhorn},
  {Krebs}, {Jhabvala}, {Gezari}, {Fixsen}, {Flores}, {Shakoorzadeh}, {Jungo},
  {Hakun}, {Workman}, {Karpati}, {Kichak}, {Whitley}, {Mann}, {Tollestrup},
  {Eisenhardt}, {Stern}, {Gorjian}, {Bhattacharya}, {Carey}, {Nelson},
  {Glaccum}, {Lacy}, {Lowrance}, {Laine}, {Reach}, {Stauffer}, {Surace},
  {Wilson}, {Wright}, {Hoffman}, {Domingo}, \& {Cohen}}]{irac}
{Fazio}, G.~G., {Hora}, J.~L., {Allen}, L.~E., {et~al.} 2004, \apjs, 154, 10

\bibitem[{{Figer} {et~al.}(2004){Figer}, {Najarro}, \& {Kudritzki}}]{figer2004}
{Figer}, D.~F., {Najarro}, F., \& {Kudritzki}, R.~P. 2004, \apjl, 610, L109

\bibitem[{{Figer} {et~al.}(1998){Figer}, {Najarro}, {Morris}, {McLean},
  {Geballe}, {Ghez}, \& {Langer}}]{figer1998}
{Figer}, D.~F., {Najarro}, F., {Morris}, M., {et~al.} 1998, \apj, 506, 384

\bibitem[{{Foellmi} {et~al.}(2008){Foellmi}, {Koenigsberger}, {Georgiev},
  {Toledano}, {Marchenko}, {Massey}, {Dall}, {Moffat}, {Morrell}, {Corcoran},
  {Kaufer}, {Naz{\'e}}, {Pittard}, {St-Louis}, {Fullerton}, {Massa}, \&
  {Pollock}}]{foellmi2008}
{Foellmi}, C., {Koenigsberger}, G., {Georgiev}, L., {et~al.} 2008, \rmxaa, 44,
  3

\bibitem[{{Freudling} {et~al.}(2013){Freudling}, {Romaniello}, {Bramich},
  {Ballester}, {Forchi}, {Garc{\'{\i}}a-Dabl{\'o}}, {Moehler}, \&
  {Neeser}}]{reflex}
{Freudling}, W., {Romaniello}, M., {Bramich}, D.~M., {et~al.} 2013, \aap, 559,
  A96

\bibitem[{{Fryer} {et~al.}(2001){Fryer}, {Woosley}, \& {Heger}}]{fryer2001}
{Fryer}, C.~L., {Woosley}, S.~E., \& {Heger}, A. 2001, \apj, 550, 372

\bibitem[{{Gal-Yam} \& {Leonard}(2009)}]{galyam2009}
{Gal-Yam}, A. \& {Leonard}, D.~C. 2009, \nat, 458, 865

\bibitem[{{Georgiev} {et~al.}(2011){Georgiev}, {Koenigsberger}, {Hillier},
  {Morrell}, {Barb{\'a}}, \& {Gamen}}]{georgiev2011}
{Georgiev}, L., {Koenigsberger}, G., {Hillier}, D.~J., {et~al.} 2011, \aj, 142,
  191

\bibitem[{{Girardi} {et~al.}(2005){Girardi}, {Groenewegen}, {Hatziminaoglou},
  \& {da Costa}}]{trilegal}
{Girardi}, L., {Groenewegen}, M.~A.~T., {Hatziminaoglou}, E., \& {da Costa}, L.
  2005, \aap, 436, 895

\bibitem[{{Groh} {et~al.}(2009){Groh}, {Damineli}, {Hillier}, {Barb{\'a}},
  {Fern{\'a}ndez-Laj{\'u}s}, {Gamen}, {Mois{\'e}s}, {Solivella}, \&
  {Teodoro}}]{groh2009}
{Groh}, J.~H., {Damineli}, A., {Hillier}, D.~J., {et~al.} 2009, \apjl, 705, L25

\bibitem[{{Groh} {et~al.}(2013){Groh}, {Meynet}, \& {Ekstr{\"o}m}}]{groh2013}
{Groh}, J.~H., {Meynet}, G., \& {Ekstr{\"o}m}, S. 2013, \aap, 550, L7

\bibitem[{{Hendry} {et~al.}(2008){Hendry}, {Smartt}, {Skillman}, {Evans},
  {Trundle}, {Lennon}, {Crowther}, \& {Hunter}}]{hendry2008}
{Hendry}, M.~A., {Smartt}, S.~J., {Skillman}, E.~D., {et~al.} 2008, \mnras,
  388, 1127

\bibitem[{{Humphreys} \& {Davidson}(1979)}]{hd1979}
{Humphreys}, R.~M. \& {Davidson}, K. 1979, \apj, 232, 409

\bibitem[{{Humphreys} \& {Davidson}(1994)}]{HD1994}
{Humphreys}, R.~M. \& {Davidson}, K. 1994, \pasp, 106, 1025

\bibitem[{{Humphreys} {et~al.}(2014){Humphreys}, {Davidson}, {Gordon}, {Weis},
  {Burggraf}, {Bomans}, \& {Martin}}]{HD2014}
{Humphreys}, R.~M., {Davidson}, K., {Gordon}, M.~S., {et~al.} 2014, \apjl, 782,
  L21

\bibitem[{{Hutsemekers} {et~al.}(1994){Hutsemekers}, {van Drom}, {Gosset}, \&
  {Melnick}}]{hut1994}
{Hutsemekers}, D., {van Drom}, E., {Gosset}, E., \& {Melnick}, J. 1994, \aap,
  290, 906

\bibitem[{{Kaib} \& {Raymond}(2014)}]{kaib2014}
{Kaib}, N.~A. \& {Raymond}, S.~N. 2014, \apj, 782, 60

\bibitem[{{Kashi}(2010)}]{kashi2010}
{Kashi}, A. 2010, \mnras, 405, 1924

\bibitem[{{Kashi} \& {Soker}(2010)}]{kashi2010b}
{Kashi}, A. \& {Soker}, N. 2010, \apj, 723, 602

\bibitem[{{Kotak} \& {Vink}(2006)}]{kotak2006}
{Kotak}, R. \& {Vink}, J.~S. 2006, \aap, 460, L5

\bibitem[{{Krumholz}(2012)}]{krum2012}
{Krumholz}, M.~R. 2012, in Astronomical Society of the Pacific Conference
  Series, Vol. 464, Circumstellar Dynamics at High Resolution, ed. A.~C.
  {Carciofi} \& T.~{Rivinius}, 339

\bibitem[{{Krumholz}(2015)}]{krum2015}
{Krumholz}, M.~R. 2015, in Astrophysics and Space Science Library, Vol. 412,
  Astrophysics and Space Science Library, ed. J.~S. {Vink}, 43

\bibitem[{{Krumholz} {et~al.}(2009){Krumholz}, {Klein}, {McKee}, {Offner}, \&
  {Cunningham}}]{krum2009}
{Krumholz}, M.~R., {Klein}, R.~I., {McKee}, C.~F., {Offner}, S.~S.~R., \&
  {Cunningham}, A.~J. 2009, Science, 323, 754

\bibitem[{{Lagage} {et~al.}(2004){Lagage}, {Pel}, {Authier}, {Belorgey},
  {Claret}, {Doucet}, {Dubreuil}, {Durand}, {Elswijk}, {Girardot}, {K{\"a}ufl},
  {Kroes}, {Lortholary}, {Lussignol}, {Marchesi}, {Pantin}, {Peletier},
  {Pirard}, {Pragt}, {Rio}, {Schoenmaker}, {Siebenmorgen}, {Silber}, {Smette},
  {Sterzik}, \& {Veyssiere}}]{visir}
{Lagage}, P.~O., {Pel}, J.~W., {Authier}, M., {et~al.} 2004, The Messenger,
  117, 12

\bibitem[{{Larsson} {et~al.}(2011){Larsson}, {Fransson}, {{\"O}stlin},
  {Gr{\"o}ningsson}, {Jerkstrand}, {Kozma}, {Sollerman}, {Challis}, {Kirshner},
  {Chevalier}, {Heng}, {McCray}, {Suntzeff}, {Bouchet}, {Crotts}, {Danziger},
  {Dwek}, {France}, {Garnavich}, {Lawrence}, {Leibundgut}, {Lundqvist},
  {Panagia}, {Pun}, {Smith}, {Sonneborn}, {Wang}, \& {Wheeler}}]{larsson2011}
{Larsson}, J., {Fransson}, C., {{\"O}stlin}, G., {et~al.} 2011, \nat, 474, 484

\bibitem[{{Le Bouquin} {et~al.}(2011){Le Bouquin}, {Berger}, {Lazareff},
  {Zins}, {Haguenauer}, {Jocou}, {Kern}, {Millan-Gabet}, {Traub}, {Absil},
  {Augereau}, {Benisty}, {Blind}, {Bonfils}, {Bourget}, {Delboulbe},
  {Feautrier}, {Germain}, {Gitton}, {Gillier}, {Kiekebusch}, {Kluska},
  {Knudstrup}, {Labeye}, {Lizon}, {Monin}, {Magnard}, {Malbet}, {Maurel},
  {M{\'e}nard}, {Micallef}, {Michaud}, {Montagnier}, {Morel}, {Moulin},
  {Perraut}, {Popovic}, {Rabou}, {Rochat}, {Rojas}, {Roussel}, {Roux},
  {Stadler}, {Stefl}, {Tatulli}, \& {Ventura}}]{pionier}
{Le Bouquin}, J.-B., {Berger}, J.-P., {Lazareff}, B., {et~al.} 2011, \aap, 535,
  A67

\bibitem[{{Lenzen} {et~al.}(2003){Lenzen}, {Hartung}, {Brandner}, {Finger},
  {Hubin}, {Lacombe}, {Lagrange}, {Lehnert}, {Moorwood}, \& {Mouillet}}]{naco1}
{Lenzen}, R., {Hartung}, M., {Brandner}, W., {et~al.} 2003, in Society of
  Photo-Optical Instrumentation Engineers (SPIE) Conference Series, Vol. 4841,
  Instrument Design and Performance for Optical/Infrared Ground-based
  Telescopes, ed. M.~{Iye} \& A.~F.~M. {Moorwood}, 944--952

\bibitem[{{L{\'e}pine} \& {Bongiorno}(2007)}]{lepine2007}
{L{\'e}pine}, S. \& {Bongiorno}, B. 2007, \aj, 133, 889

\bibitem[{{Lobel} {et~al.}(2013){Lobel}, {Groh}, {Martayan}, {Fr{\'e}mat},
  {Torres Dozinel}, {Raskin}, {Van Winckel}, {Prins}, {Pessemier}, {Waelkens},
  {Hensberge}, {Dumortier}, {Jorissen}, {Van Eck}, \& {Lehmann}}]{lobel2013}
{Lobel}, A., {Groh}, J.~H., {Martayan}, C., {et~al.} 2013, \aap, 559, A16

\bibitem[{{Longhitano} \& {Binggeli}(2010)}]{long2010}
{Longhitano}, M. \& {Binggeli}, B. 2010, \aap, 509, A46

\bibitem[{{Maeder} \& {Desjacques}(2001)}]{md2001}
{Maeder}, A. \& {Desjacques}, V. 2001, \aap, 372, L9

\bibitem[{{Marchenko} \& {Moffat}(2007)}]{marchenko2007}
{Marchenko}, S.~V. \& {Moffat}, A.~F.~J. 2007, in Astronomical Society of the
  Pacific Conference Series, Vol. 367, Massive Stars in Interactive Binaries,
  ed. N.~{St.-Louis} \& A.~F.~J. {Moffat}, 213

\bibitem[{{Marston} \& {McCollum}(2008)}]{marston2008}
{Marston}, A.~P. \& {McCollum}, B. 2008, \aap, 477, 193

\bibitem[{{Martayan} {et~al.}(2012){Martayan}, {Lobel}, {Baade}, {Blomme},
  {Fr{\'e}mat}, {LeBouquin}, {Selman}, {Girard}, {M{\'e}rand}, {Montagnier},
  {Patru}, {Mawet}, {Martins}, {Rivinius}, {{\v S}tefl}, {Zorec}, {Semaan},
  {Mehner}, {Kervella}, {Sana}, \& {Sch{\"o}del}}]{martayan2012}
{Martayan}, C., {Lobel}, A., {Baade}, D., {et~al.} 2012, in Astronomical
  Society of the Pacific Conference Series, Vol. 464, Circumstellar Dynamics at
  High Resolution, ed. A.~C. {Carciofi} \& T.~{Rivinius}, 293

\bibitem[{{Mason} {et~al.}(2009){Mason}, {Hartkopf}, {Gies}, {Henry}, \&
  {Helsel}}]{mason2009}
{Mason}, B.~D., {Hartkopf}, W.~I., {Gies}, D.~R., {Henry}, T.~J., \& {Helsel},
  J.~W. 2009, \aj, 137, 3358

\bibitem[{{Miroshnichenko} {et~al.}(2005){Miroshnichenko}, {Bjorkman},
  {Grosso}, {Levato}, {Grankin}, {Rudy}, {Lynch}, {Mazuk}, \&
  {Puetter}}]{miro2005}
{Miroshnichenko}, A.~S., {Bjorkman}, K.~S., {Grosso}, M., {et~al.} 2005,
  \mnras, 364, 335

\bibitem[{{Miroshnichenko} {et~al.}(1998){Miroshnichenko}, {Fremat},
  {Houziaux}, {Andrillat}, {Chentsov}, \& {Klochkova}}]{miro1998}
{Miroshnichenko}, A.~S., {Fremat}, Y., {Houziaux}, L., {et~al.} 1998, \aaps,
  131, 469

\bibitem[{{Morris} \& {Podsiadlowski}(2009)}]{morris2009}
{Morris}, T. \& {Podsiadlowski}, P. 2009, \mnras, 399, 515

\bibitem[{{Naz{\'e}} {et~al.}(2012){Naz{\'e}}, {Rauw}, \&
  {Hutsem{\'e}kers}}]{naze2012}
{Naz{\'e}}, Y., {Rauw}, G., \& {Hutsem{\'e}kers}, D. 2012, \aap, 538, A47

\bibitem[{{Nota} {et~al.}(1996){Nota}, {Pasquali}, {Clampin}, {Pollacco},
  {Scuderi}, \& {Livio}}]{nota1996}
{Nota}, A., {Pasquali}, A., {Clampin}, M., {et~al.} 1996, \apj, 473, 946

\bibitem[{{O'Hara} {et~al.}(2003){O'Hara}, {Meixner}, {Speck}, {Ueta}, \&
  {Bobrowsky}}]{ohara2003}
{O'Hara}, T.~B., {Meixner}, M., {Speck}, A.~K., {Ueta}, T., \& {Bobrowsky}, M.
  2003, \apj, 598, 1255

\bibitem[{{Oskinova} {et~al.}(2013){Oskinova}, {Steinke}, {Hamann}, {Sander},
  {Todt}, \& {Liermann}}]{os2013}
{Oskinova}, L.~M., {Steinke}, M., {Hamann}, W.-R., {et~al.} 2013, \mnras, 436,
  3357

\bibitem[{{Oudmaijer} \& {Parr}(2010)}]{oud2010}
{Oudmaijer}, R.~D. \& {Parr}, A.~M. 2010, \mnras, 405, 2439

\bibitem[{{Owocki}(2011)}]{owocki2011}
{Owocki}, S. 2011, Bulletin de la Societe Royale des Sciences de Liege, 80, 16

\bibitem[{{Pasquali} {et~al.}(2002){Pasquali}, {Nota}, {Smith}, {Akiyama},
  {Messineo}, \& {Clampin}}]{pasquali2002}
{Pasquali}, A., {Nota}, A., {Smith}, L.~J., {et~al.} 2002, \aj, 124, 1625

\bibitem[{{Petrov} {et~al.}(2007){Petrov}, {Malbet}, {Weigelt}, {Antonelli},
  {Beckmann}, {Bresson}, {Chelli}, {Dugu{\'e}}, {Duvert}, {Gennari},
  {Gl{\"u}ck}, {Kern}, {Lagarde}, {Le Coarer}, {Lisi}, {Millour}, {Perraut},
  {Puget}, {Rantakyr{\"o}}, {Robbe-Dubois}, {Roussel}, {Salinari}, {Tatulli},
  {Zins}, {Accardo}, {Acke}, {Agabi}, {Altariba}, {Arezki}, {Aristidi},
  {Baffa}, {Behrend}, {Bl{\"o}cker}, {Bonhomme}, {Busoni}, {Cassaing},
  {Clausse}, {Colin}, {Connot}, {Delboulb{\'e}}, {Domiciano de Souza},
  {Driebe}, {Feautrier}, {Ferruzzi}, {Forveille}, {Fossat}, {Foy},
  {Fraix-Burnet}, {Gallardo}, {Giani}, {Gil}, {Glentzlin}, {Heiden},
  {Heininger}, {Hernandez Utrera}, {Hofmann}, {Kamm}, {Kiekebusch}, {Kraus},
  {Le Contel}, {Le Contel}, {Lesourd}, {Lopez}, {Lopez}, {Magnard}, {Marconi},
  {Mars}, {Martinot-Lagarde}, {Mathias}, {M{\`e}ge}, {Monin}, {Mouillet},
  {Mourard}, {Nussbaum}, {Ohnaka}, {Pacheco}, {Perrier}, {Rabbia}, {Rebattu},
  {Reynaud}, {Richichi}, {Robini}, {Sacchettini}, {Schertl}, {Sch{\"o}ller},
  {Solscheid}, {Spang}, {Stee}, {Stefanini}, {Tallon}, {Tallon-Bosc}, {Tasso},
  {Testi}, {Vakili}, {von der L{\"u}he}, {Valtier}, {Vannier}, \&
  {Ventura}}]{amber}
{Petrov}, R.~G., {Malbet}, F., {Weigelt}, G., {et~al.} 2007, \aap, 464, 1

\bibitem[{{Pilbratt} {et~al.}(2010){Pilbratt}, {Riedinger}, {Passvogel},
  {Crone}, {Doyle}, {Gageur}, {Heras}, {Jewell}, {Metcalfe}, {Ott}, \&
  {Schmidt}}]{herschell2010}
{Pilbratt}, G.~L., {Riedinger}, J.~R., {Passvogel}, T., {et~al.} 2010, \aap,
  518, L1

\bibitem[{{Rajan}(2010)}]{wfc}
{Rajan}, A. e.~a. 2010, in WFC3 Data Handbook Version 2.1, Baltimore: STScI

\bibitem[{{Rauw} {et~al.}(2012){Rauw}, {Sana}, {Spano}, {Gosset}, {Mahy}, {De
  Becker}, \& {Eenens}}]{rauw2012}
{Rauw}, G., {Sana}, H., {Spano}, M., {et~al.} 2012, \aap, 542, A95

\bibitem[{{Reid}(1993)}]{reid1993}
{Reid}, M.~J. 1993, \araa, 31, 345

\bibitem[{{Rieke} {et~al.}(2004){Rieke}, {Young}, {Engelbracht}, {Kelly},
  {Low}, {Haller}, {Beeman}, {Gordon}, {Stansberry}, {Misselt}, {Cadien},
  {Morrison}, {Rivlis}, {Latter}, {Noriega-Crespo}, {Padgett}, {Stapelfeldt},
  {Hines}, {Egami}, {Muzerolle}, {Alonso-Herrero}, {Blaylock}, {Dole}, {Hinz},
  {Le Floc'h}, {Papovich}, {P{\'e}rez-Gonz{\'a}lez}, {Smith}, {Su}, {Bennett},
  {Frayer}, {Henderson}, {Lu}, {Masci}, {Pesenson}, {Rebull}, {Rho}, {Keene},
  {Stolovy}, {Wachter}, {Wheaton}, {Werner}, \& {Richards}}]{mips}
{Rieke}, G.~H., {Young}, E.~T., {Engelbracht}, C.~W., {et~al.} 2004, \apjs,
  154, 25

\bibitem[{{Rivinius} {et~al.}(2015){Rivinius}, {Boffin}, {de Wit}, {Mehner},
  {Martayan}, {Guieu}, \& {Le Bouquin}}]{rivinius2014}
{Rivinius}, T., {Boffin}, H.~M.~J., {de Wit}, W.~J., {et~al.} 2015, in IAU
  Symposium, Vol. 307, IAU Symposium, 295--296

\bibitem[{{Roming} {et~al.}(2005){Roming}, {Kennedy}, {Mason}, {Nousek}, {Ahr},
  {Bingham}, {Broos}, {Carter}, {Hancock}, {Huckle}, {Hunsberger}, {Kawakami},
  {Killough}, {Koch}, {McLelland}, {Smith}, {Smith}, {Soto}, {Boyd},
  {Breeveld}, {Holland}, {Ivanushkina}, {Pryzby}, {Still}, \&
  {Stock}}]{swiftuvot2005}
{Roming}, P.~W.~A., {Kennedy}, T.~E., {Mason}, K.~O., {et~al.} 2005, \ssr, 120,
  95

\bibitem[{{Rousset} {et~al.}(2003){Rousset}, {Lacombe}, {Puget}, {Hubin},
  {Gendron}, {Fusco}, {Arsenault}, {Charton}, {Feautrier}, {Gigan}, {Kern},
  {Lagrange}, {Madec}, {Mouillet}, {Rabaud}, {Rabou}, {Stadler}, \&
  {Zins}}]{naco2}
{Rousset}, G., {Lacombe}, F., {Puget}, P., {et~al.} 2003, in Society of
  Photo-Optical Instrumentation Engineers (SPIE) Conference Series, Vol. 4839,
  Adaptive Optical System Technologies II, ed. P.~L. {Wizinowich} \&
  D.~{Bonaccini}, 140--149

\bibitem[{{Sana} {et~al.}(2012){Sana}, {de Mink}, {de Koter}, {Langer},
  {Evans}, {Gieles}, {Gosset}, {Izzard}, {Le Bouquin}, \&
  {Schneider}}]{sana2012}
{Sana}, H., {de Mink}, S.~E., {de Koter}, A., {et~al.} 2012, Science, 337, 444

\bibitem[{{Sana} {et~al.}(2008){Sana}, {Gosset}, {Naz{\'e}}, {Rauw}, \&
  {Linder}}]{sana2008}
{Sana}, H., {Gosset}, E., {Naz{\'e}}, Y., {Rauw}, G., \& {Linder}, N. 2008,
  \mnras, 386, 447

\bibitem[{{Sana} \& {Le Bouquin}(2010)}]{sanascan2010}
{Sana}, H. \& {Le Bouquin}, J.-B. 2010, in Revista Mexicana de Astronomia y
  Astrofisica Conference Series, Vol.~38, 27--29

\bibitem[{{Sana} {et~al.}(2014){Sana}, {Le Bouquin}, {Lacour}, {Berger},
  {Duvert}, {Gauchet}, {Norris}, {Olofsson}, {Pickel}, {Zins}, {Absil}, {de
  Koter}, {Kratter}, {Schnurr}, \& {Zinnecker}}]{sana2014}
{Sana}, H., {Le Bouquin}, J.-B., {Lacour}, S., {et~al.} 2014, \apjs, 215, 15

\bibitem[{{Schaller} {et~al.}(1992){Schaller}, {Schaerer}, {Meynet}, \&
  {Maeder}}]{schaller1992}
{Schaller}, G., {Schaerer}, D., {Meynet}, G., \& {Maeder}, A. 1992, \aaps, 96,
  269

\bibitem[{{Shaya} \& {Olling}(2011)}]{shaya2011}
{Shaya}, E.~J. \& {Olling}, R.~P. 2011, \apjs, 192, 2

\bibitem[{{Skrutskie} {et~al.}(2006){Skrutskie}, {Cutri}, {Stiening},
  {Weinberg}, {Schneider}, {Carpenter}, {Beichman}, {Capps}, {Chester},
  {Elias}, {Huchra}, {Liebert}, {Lonsdale}, {Monet}, {Price}, {Seitzer},
  {Jarrett}, {Kirkpatrick}, {Gizis}, {Howard}, {Evans}, {Fowler}, {Fullmer},
  {Hurt}, {Light}, {Kopan}, {Marsh}, {McCallon}, {Tam}, {Van Dyk}, \&
  {Wheelock}}]{2mass}
{Skrutskie}, M.~F., {Cutri}, R.~M., {Stiening}, R., {et~al.} 2006, \aj, 131,
  1163

\bibitem[{{Smith}(2007)}]{smith2007}
{Smith}, N. 2007, \aj, 133, 1034

\bibitem[{{Smith} {et~al.}(2013){Smith}, {Arnett}, {Bally}, {Ginsburg}, \&
  {Filippenko}}]{smith2013}
{Smith}, N., {Arnett}, W.~D., {Bally}, J., {Ginsburg}, A., \& {Filippenko},
  A.~V. 2013, \mnras, 429, 1324

\bibitem[{{Smith} {et~al.}(2007){Smith}, {Bally}, \& {Walawender}}]{smith2007b}
{Smith}, N., {Bally}, J., \& {Walawender}, J. 2007, \aj, 134, 846

\bibitem[{{Smith} {et~al.}(2011){Smith}, {Li}, {Miller}, {Silverman},
  {Filippenko}, {Cuillandre}, {Cooper}, {Matheson}, \& {Van Dyk}}]{smith2011}
{Smith}, N., {Li}, W., {Miller}, A.~A., {et~al.} 2011, \apj, 732, 63

\bibitem[{{Smith} \& {Tombleson}(2015)}]{smith2015}
{Smith}, N. \& {Tombleson}, R. 2015, \mnras, 447, 598

\bibitem[{{Soker}(2004)}]{soker2004}
{Soker}, N. 2004, \apj, 612, 1060

\bibitem[{{Sung} {et~al.}(2000){Sung}, {Chun}, \& {Bessell}}]{sung2000}
{Sung}, H., {Chun}, M.-Y., \& {Bessell}, M.~S. 2000, \aj, 120, 333

\bibitem[{{Tan} {et~al.}(2014){Tan}, {Beltran}, {Caselli}, {Fontani}, {Fuente},
  {Krumholz}, {McKee}, \& {Stolte}}]{tan2014}
{Tan}, J.~C., {Beltran}, M.~T., {Caselli}, P., {et~al.} 2014, ArXiv e-prints

\bibitem[{{Thompson}(1994)}]{nicmos}
{Thompson}, R.~I. 1994, in Society of Photo-Optical Instrumentation Engineers
  (SPIE) Conference Series, Vol. 2198, Instrumentation in Astronomy VIII, ed.
  D.~L. {Crawford} \& E.~R. {Craine}, 1202--1213

\bibitem[{{Trundle} {et~al.}(2008){Trundle}, {Kotak}, {Vink}, \&
  {Meikle}}]{trundle2008}
{Trundle}, C., {Kotak}, R., {Vink}, J.~S., \& {Meikle}, W.~P.~S. 2008, \aap,
  483, L47

\bibitem[{{Tylenda} {et~al.}(2003){Tylenda}, {Si{\'o}dmiak}, {G{\'o}rny},
  {Corradi}, \& {Schwarz}}]{PN2003}
{Tylenda}, R., {Si{\'o}dmiak}, N., {G{\'o}rny}, S.~K., {Corradi}, R.~L.~M., \&
  {Schwarz}, H.~E. 2003, \aap, 405, 627

\bibitem[{{Ubeda}(2014)}]{acs}
{Ubeda}, L. e.~a. 2014, in ACS Instrument Handbook Version 13.0, Baltimore:
  STScI

\bibitem[{{Van Dyk} \& {Matheson}(2012)}]{vd2012}
{Van Dyk}, S.~D. \& {Matheson}, T. 2012, \apj, 746, 179

\bibitem[{{van Genderen}(2001)}]{van2001}
{van Genderen}, A.~M. 2001, \aap, 366, 508

\bibitem[{{van Marle} {et~al.}(2008){van Marle}, {Owocki}, \&
  {Shaviv}}]{vm2008}
{van Marle}, A.~J., {Owocki}, S.~P., \& {Shaviv}, N.~J. 2008, \mnras, 389, 1353

\bibitem[{{Vernet} {et~al.}(2011){Vernet}, {Dekker}, {D'Odorico}, {Kaper},
  {Kjaergaard}, {Hammer}, {Randich}, {Zerbi}, {Groot}, {Hjorth}, {Guinouard},
  {Navarro}, {Adolfse}, {Albers}, {Amans}, {Andersen}, {Andersen}, {Binetruy},
  {Bristow}, {Castillo}, {Chemla}, {Christensen}, {Conconi}, {Conzelmann},
  {Dam}, {de Caprio}, {de Ugarte Postigo}, {Delabre}, {di Marcantonio},
  {Downing}, {Elswijk}, {Finger}, {Fischer}, {Flores}, {Fran{\c c}ois},
  {Goldoni}, {Guglielmi}, {Haigron}, {Hanenburg}, {Hendriks}, {Horrobin},
  {Horville}, {Jessen}, {Kerber}, {Kern}, {Kiekebusch}, {Kleszcz}, {Klougart},
  {Kragt}, {Larsen}, {Lizon}, {Lucuix}, {Mainieri}, {Manuputy}, {Martayan},
  {Mason}, {Mazzoleni}, {Michaelsen}, {Modigliani}, {Moehler}, {M{\o}ller},
  {Norup S{\o}rensen}, {N{\o}rregaard}, {P{\'e}roux}, {Patat}, {Pena}, {Pragt},
  {Reinero}, {Rigal}, {Riva}, {Roelfsema}, {Royer}, {Sacco}, {Santin},
  {Schoenmaker}, {Spano}, {Sweers}, {Ter Horst}, {Tintori}, {Tromp}, {van
  Dael}, {van der Vliet}, {Venema}, {Vidali}, {Vinther}, {Vola}, {Winters},
  {Wistisen}, {Wulterkens}, \& {Zacchei}}]{xshooter}
{Vernet}, J., {Dekker}, H., {D'Odorico}, S., {et~al.} 2011, \aap, 536, A105

\bibitem[{{Weis}(2011)}]{weis2011}
{Weis}, K. 2011, Bulletin de la Societe Royale des Sciences de Liege, 80, 440

\bibitem[{{Werner} {et~al.}(2004){Werner}, {Roellig}, {Low}, {Rieke}, {Rieke},
  {Hoffmann}, {Young}, {Houck}, {Brandl}, {Fazio}, {Hora}, {Gehrz}, {Helou},
  {Soifer}, {Stauffer}, {Keene}, {Eisenhardt}, {Gallagher}, {Gautier}, {Irace},
  {Lawrence}, {Simmons}, {Van Cleve}, {Jura}, {Wright}, \&
  {Cruikshank}}]{spitzer}
{Werner}, M.~W., {Roellig}, T.~L., {Low}, F.~J., {et~al.} 2004, \apjs, 154, 1

\bibitem[{{Wright} {et~al.}(2010){Wright}, {Eisenhardt}, {Mainzer}, {Ressler},
  {Cutri}, {Jarrett}, {Kirkpatrick}, {Padgett}, {McMillan}, {Skrutskie},
  {Stanford}, {Cohen}, {Walker}, {Mather}, {Leisawitz}, {Gautier}, {McLean},
  {Benford}, {Lonsdale}, {Blain}, {Mendez}, {Irace}, {Duval}, {Liu}, {Royer},
  {Heinrichsen}, {Howard}, {Shannon}, {Kendall}, {Walsh}, {Larsen}, {Cardon},
  {Schick}, {Schwalm}, {Abid}, {Fabinsky}, {Naes}, \& {Tsai}}]{WISE}
{Wright}, E.~L., {Eisenhardt}, P.~R.~M., {Mainzer}, A.~K., {et~al.} 2010, \aj,
  140, 1868

\end{thebibliography}


 \Online

\begin{appendix} 

\section{Description of the circumstellar environment }
\label{GCED2}
 
In the following sections, the infrared images of the stars with their main circumstellar environment structures are discussed.


\subsection{Pistol
Star}
\begin{figure*}[ht!]
   \centering
   \includegraphics[width=\hsize]{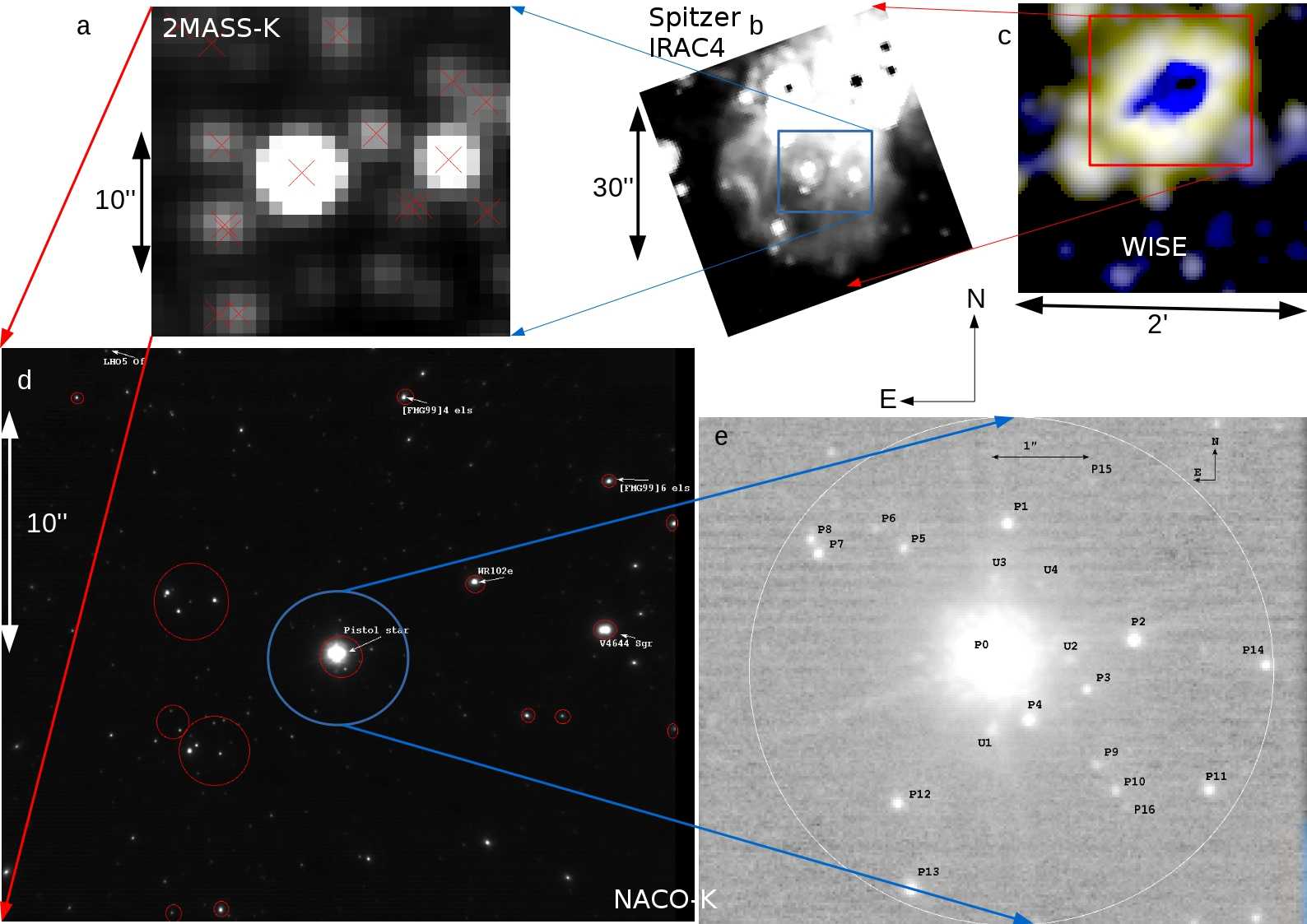}
      \caption{Multiband images of the Pistol
Star. 
      Top row: a) $28\arcsec \times 25\arcsec$ 2MASS K-band image, the red crosses correspond to 2MASS known objects; b) $1.12\arcmin \times 1.12\arcmin$ Spitzer irac4 image treated to enhance the surrounding nebula of the Pistol
Star;
      c) WISE $2\arcmin \times 2\arcmin$ ``tricolour'' image R=WISE2, G=WISE2, B=WISE1, the channels 4 and 3 are saturated and cannot be used. This image shows the large nebula surrounding Pistol
Star.
      Bottom row: d) NACO $28\arcsec \times 25\arcsec$ K image, the red circles correspond to 2MASS stars; several known stars are identified in the image. It illustrates the adaptive optics capability in the field; e) zoom ($7\arcsec \times 5\arcsec$) of image d where the surrounding stars are numbered. The white circle has a 5.5$\arcsec$ diameter. The objects labelled ``u'' are possible artefacts.
      For all images, the orientation is north-top, east-left. }
         \label{FigPS}
   \end{figure*}

In  Fig.~\ref{FigPS}, we present infrared images of the Pistol
Star and its surrounding environment including WR102e, among others.
The comparison between 2MASS and NACO images illustrates the adaptive optics capability. The WISE and Spitzer-IRAC images show the large shell ejected by the Pistol
Star. Our NACO images show the nearby stars of the Pistol
Star (with a zoom) and WR102e. 
We  also note in this image the resolved double star V4644 Sgr.

\subsection{WR102ka}
\begin{figure*}[h!]
   \centering
   \includegraphics[width=\hsize]{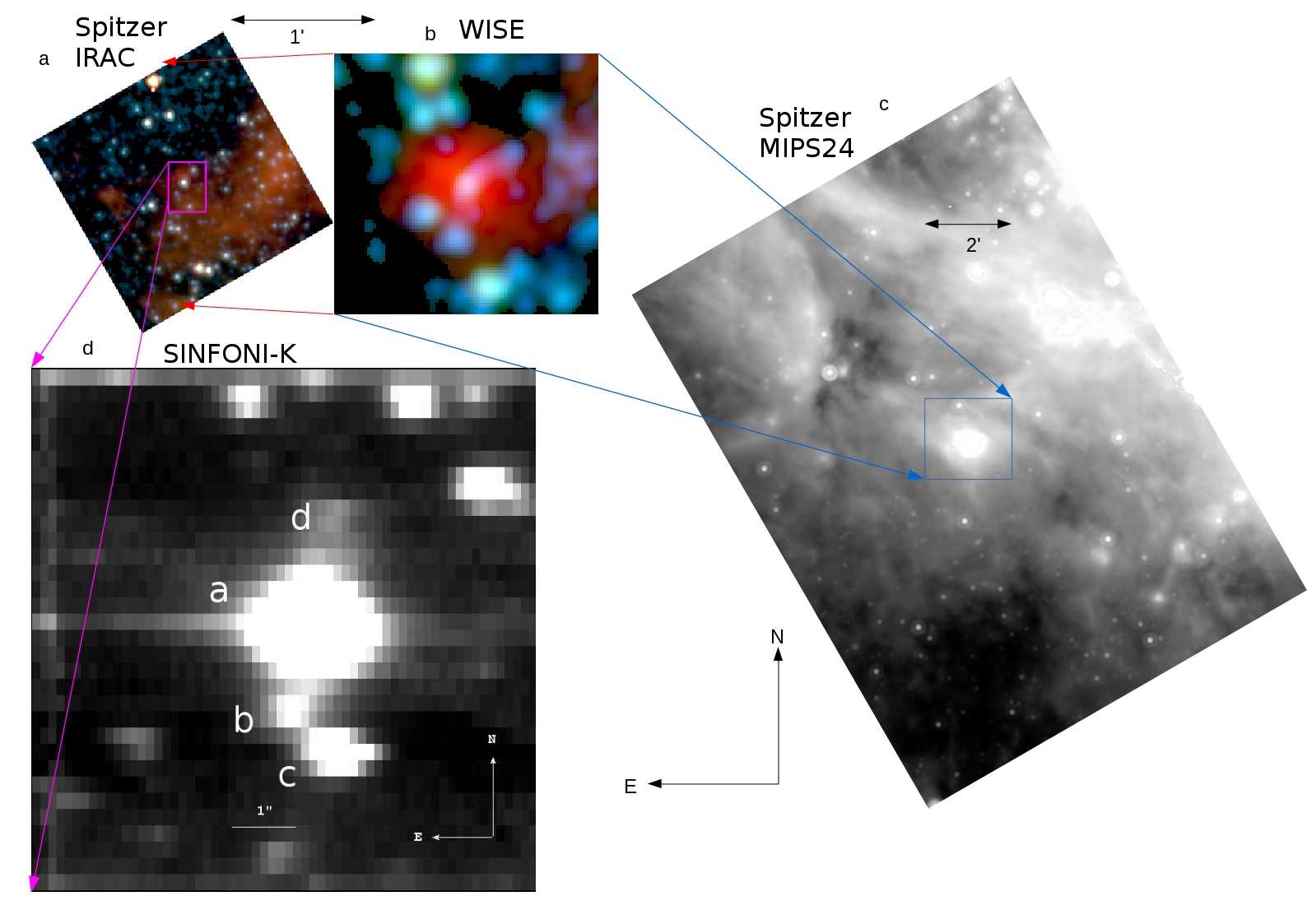}
      \caption{Multiband images of WR102ka. a) $2.75\arcmin \times 2.75\arcmin$ Spitzer tricolour image R=irac4, G=irac3, B=irac2; b) WISE $2\arcmin \times 2\arcmin$ tricolour image R=WISE3, G=WISE2, B=WISE1.
       c) Spitzer-$19\arcmin \times 19\arcmin$ MIPS-$24~\mu$m image; 
      d) reconstructed image from SINFONI in K ($8\arcsec \times 8\arcsec$). The object labelled ``d'' is possibly an artefact.
       }
         \label{Figwr102ka}
   \end{figure*}

The infrared images  in Fig.~\ref{Figwr102ka} show that the star is embedded in a complex ambient nebula, but the WISE image also shows that the star is surrounded by an almost spherical shell.
The bottom part of the figure is the reconstructed image (median) over the SINFONI cube in the K band.
See the main text of the article for the discussion of the nearby stars.

\subsection{LBV1806-20}

\begin{figure*}[h!]
   \centering
     \includegraphics[width=\hsize]{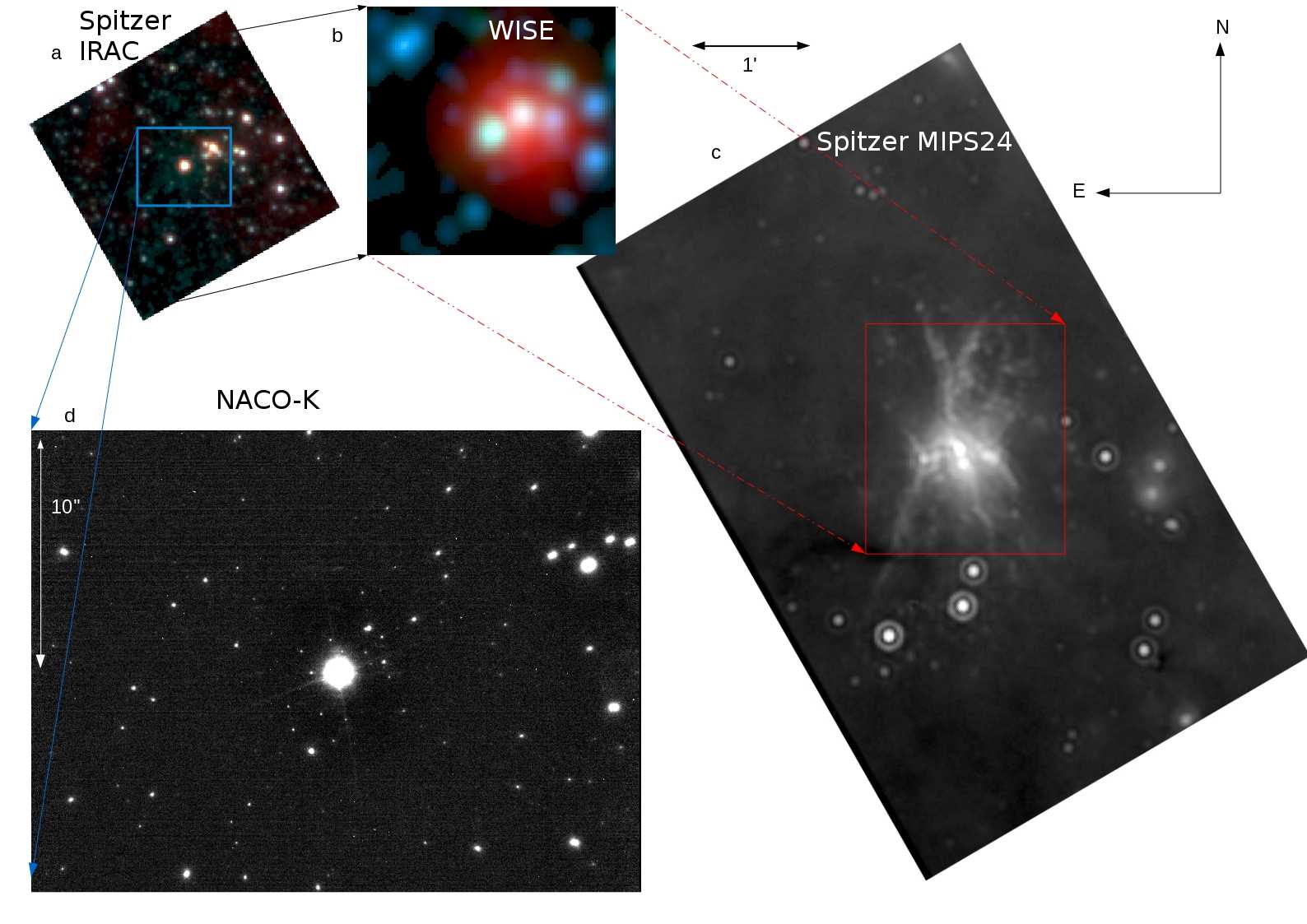}
      \caption{Multiband images of LBV1806-20. a) $2.75\arcmin \times 2.75\arcmin$ Spitzer tricolour image R=irac4, G=irac2, B=irac1; b) WISE $2\arcmin \times 2\arcmin$ tricolour image R=WISE3, G=WISE2, B=WISE1; 
      c) Spitzer-$10\arcmin \times 12\arcmin$ MIPS-$24~\mu$m image. In this image the LBV1806-20 is on the left side of the central tri-structure; d) $28\arcsec \times 21\arcsec$ NACO image in the K band.
       }
         \label{Figlbv1806}
   \end{figure*}

The Spitzer and WISE images in Fig.~\ref{Figlbv1806} show that the star LBV1806-20 lies in a complex region with massive stars and with filaments and nebulae. None of these structures is centred in LBV1806-20, but with the other two WR stars, it probably shapes this surrounding environment. 
The NACO image shows the nearby stars of LBV1806-20 discussed in the main text.

\subsection{HD168607}
\begin{figure*}[h!]
   \centering
   \includegraphics[width=\hsize]{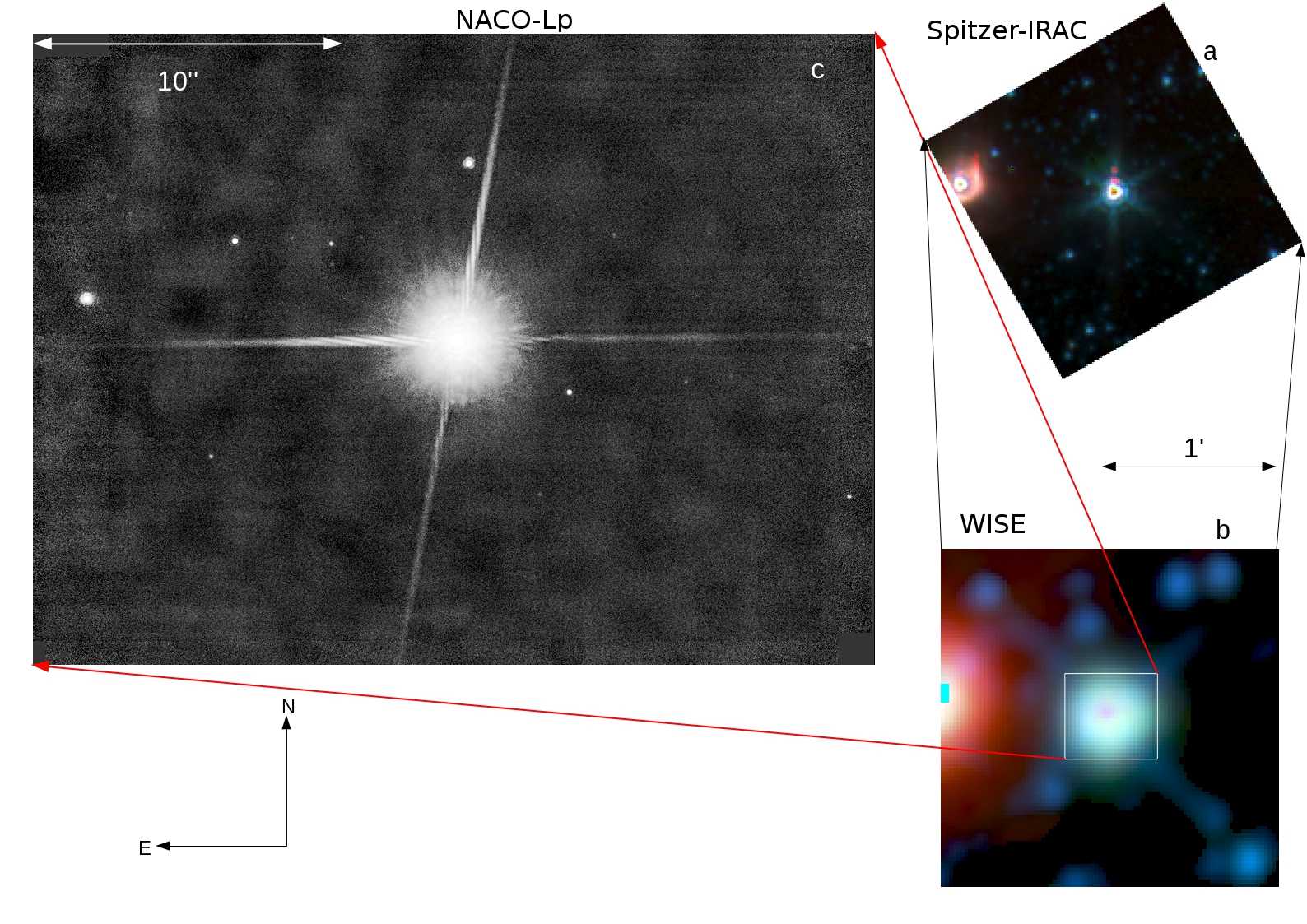}
      \caption{Multiband images of HD168607. a) $2.75\arcmin \times 2.75\arcmin$ Spitzer tricolour image R=irac4, G=irac2, B=irac1; b) WISE $2\arcmin \times 2\arcmin$ tricolour image R=WISE3, G=WISE2, B=WISE1;
       c) NACO $28\arcsec \times 21\arcsec$ Lp image. 
      }
         \label{Fighd168607}
   \end{figure*}

In Fig.~\ref{Fighd168607},  the surrounding environment of HD168607  appears bluer (warmer) than in the LBV neighbour HD168625. The shells of the latter reach the vicinity of HD168607.

\subsection{MWC930}

\begin{figure*}[h!]
   \centering
   \includegraphics[width=\hsize]{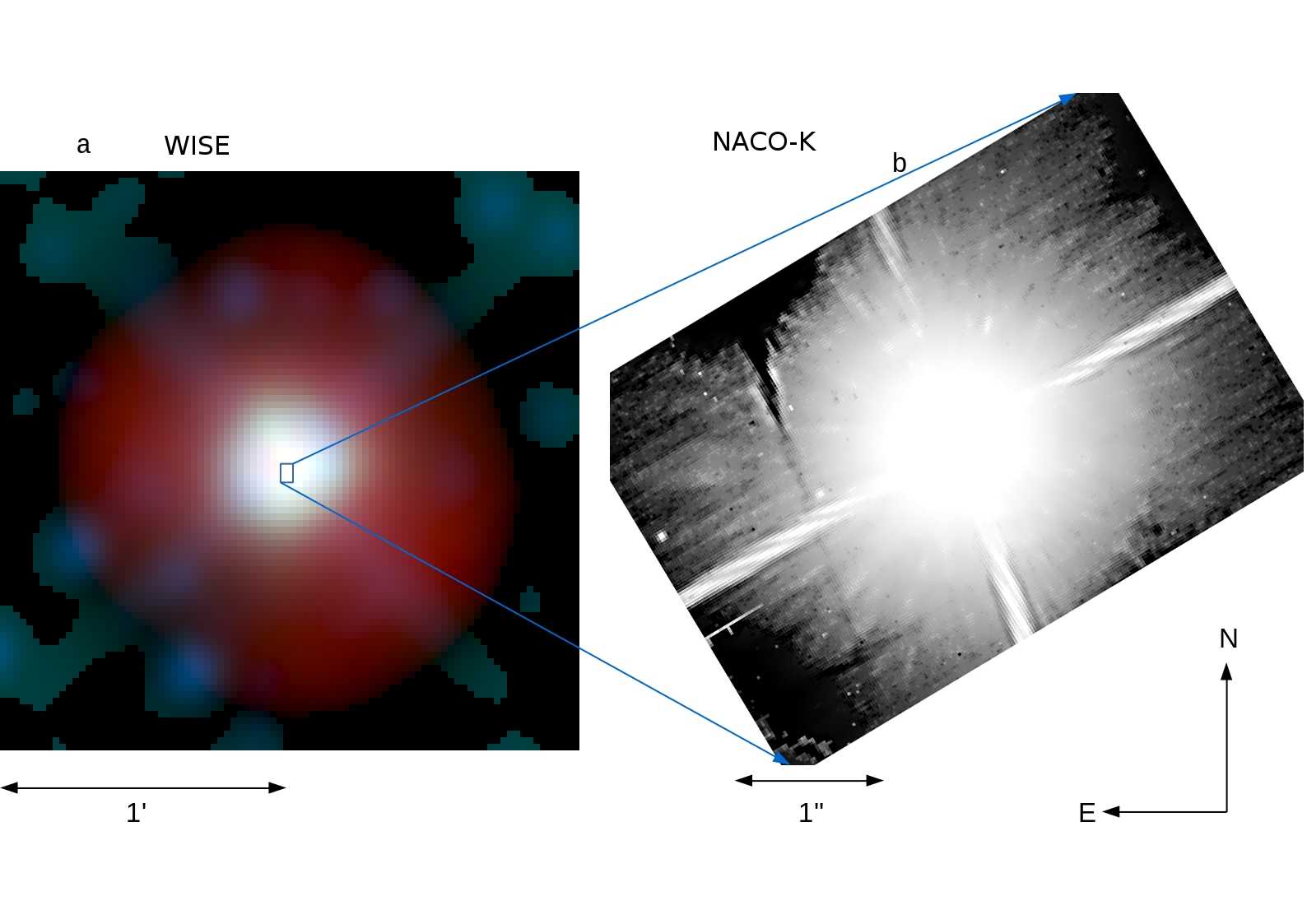}
      \caption{Multiband images of MWC930. Left: a) WISE $2\arcmin \times 2\arcmin$ tricolour image R=WISE4, G=WISE2, B=WISE1. It shows the  nebula surrounding MWC930. Right: b) NACO $7\arcsec \times 5\arcsec$ K-band image; there are several nearby objects. }
         \label{Figmwc930}
   \end{figure*}

Fig.~\ref{Figmwc930} presents the large circular shell of MWC930 visible in the WISE images.
The objects next to MWC930 seen in the NACO image are discussed in the main text.

\subsection{MWC314}  

\begin{figure*}[h!]
   \centering
   \includegraphics[width=\hsize]{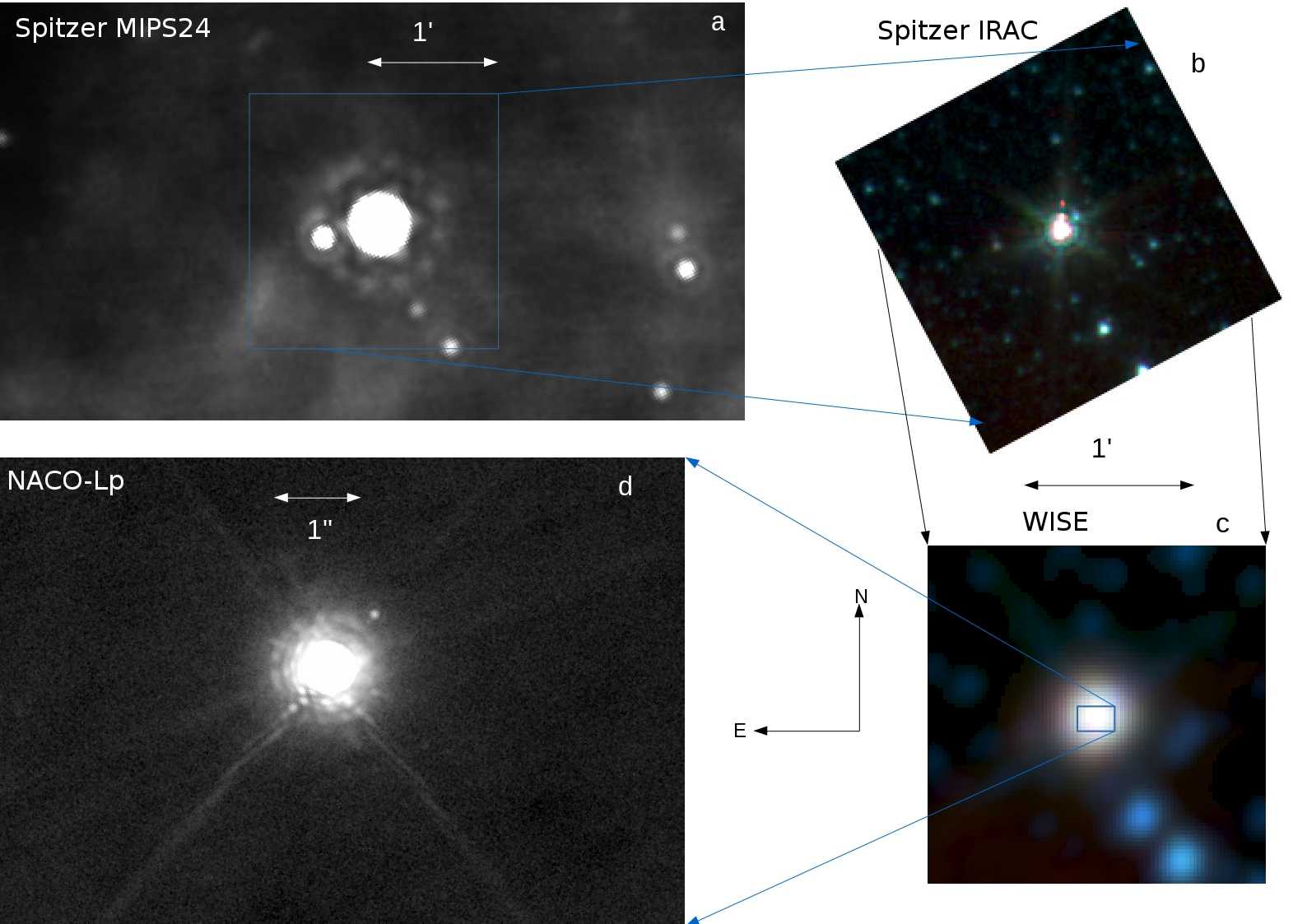}
      \caption{Multiband images of MWC314. a)  Spitzer-$6\arcmin \times 2\arcmin$ MIPS-$24~\mu$m image; 
      b) $2.75\arcmin \times 2.75\arcmin$ Spitzer tricolour image R=irac4, G=irac2, B=irac1, the blue and red spots close to the central object are artefacts due to the high flux levels of MWC314 in those channels;
      c) WISE $2\arcmin \times 2\arcmin$ tricolour image R=WISE3, G=WISE2, B=WISE1. It shows the  nebula surrounding MWC314; d) NACO $8\arcsec \times 6\arcsec$ Lp image. The nearby star is at 1.18$\arcsec$ NNW.
       }
         \label{Figmwc314}
   \end{figure*}

Fig.~\ref{Figmwc314} presents the infrared images of MWC314. The Spitzer and WISE images show the circumstellar nebula that appears less bright than in other LBVs.
In the NACO-Lp image,  the wide companion discussed in the main text can be seen.

\subsection{HD152234}
\begin{figure*}[h!]
   \centering
   \includegraphics[width=\hsize]{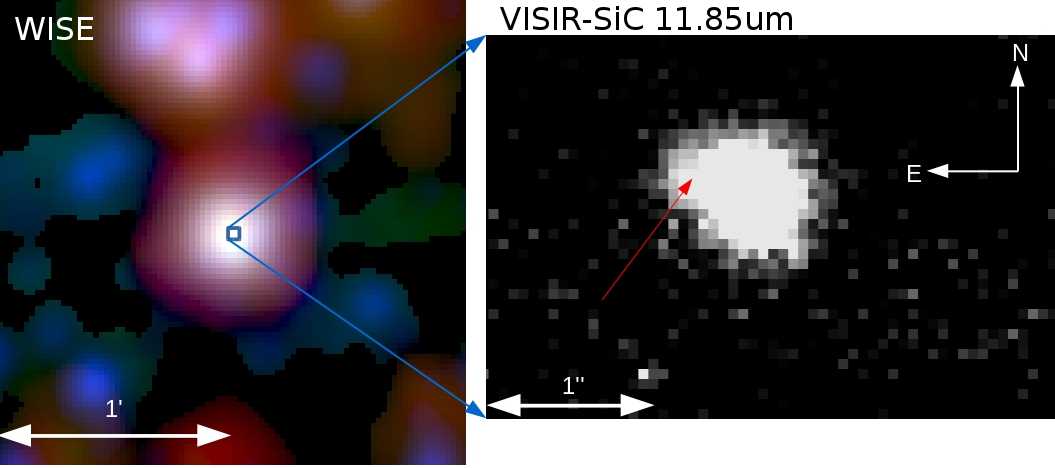}
      \caption{Multiband images of HD152234.
      Left: WISE $2\arcmin \times 2\arcmin$ tricolour image R=WISE4, G=WISE3, B=WISE1. 
      Right: VISIR-SiC $4.2\arcsec \times 2.9\arcsec$ image. There is a fainter star indicated by a red arrow at 73$\degr$ NE and at 0.5$\arcsec$ from the main object.
      }
         \label{Fighd152234}
   \end{figure*}

\cite{sana2014} indicate that HD152234 is a complex system with a close binary system called Aa and Ab that they resolved with PIONIER \citep{pionier}.
As shown in Fig.~\ref{Fighd152234}-right, we resolved  the B companion at 0.5$\arcsec$ with VISIR data. 
The nebular envelope of the system is shown in Fig.~\ref{Fighd152234}-left.

\subsection{HD164794}   

\begin{figure*}[h!]
   \centering
   \includegraphics[width=\hsize]{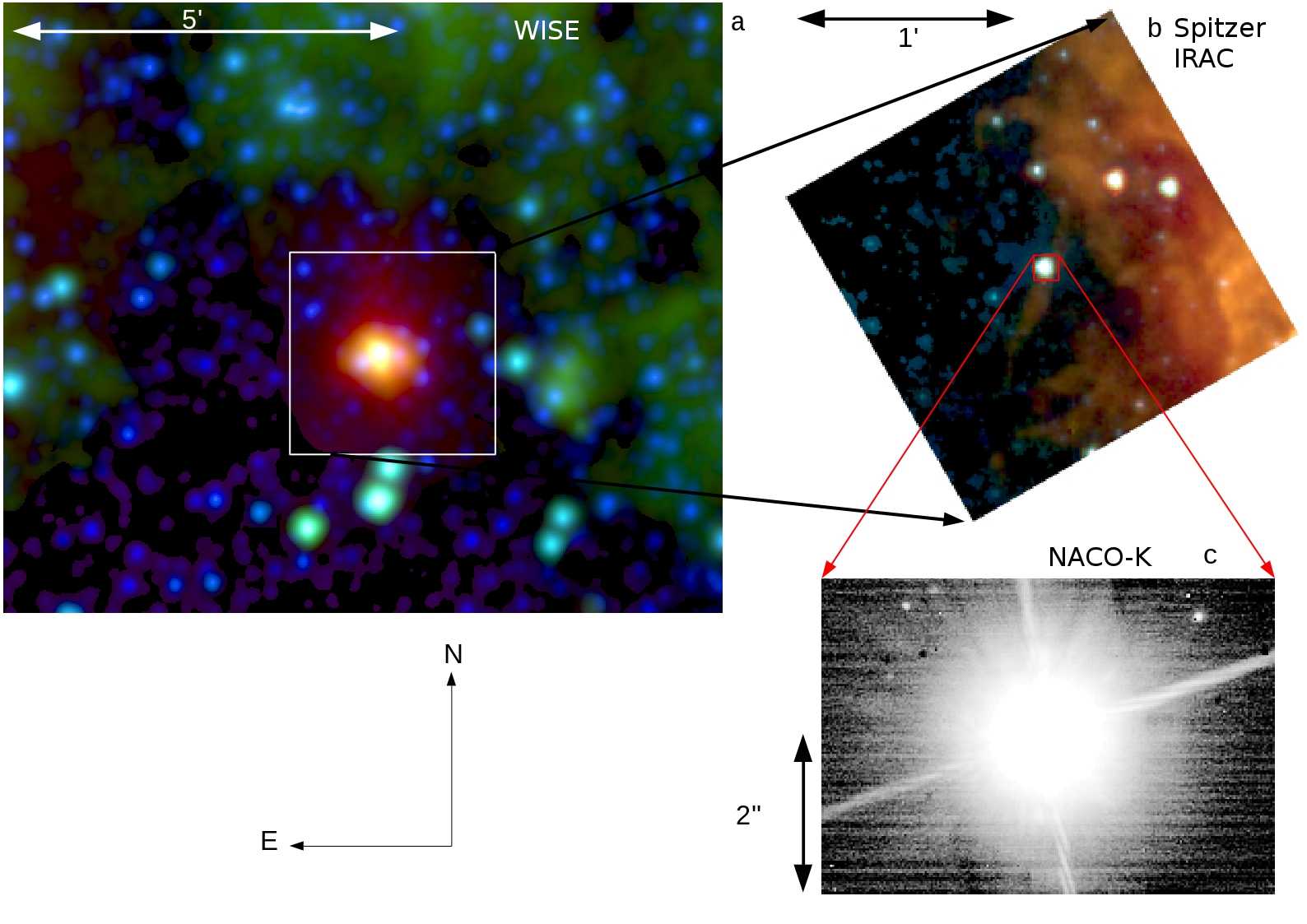}
       \caption{Multiband images of HD164794. a) WISE $10\arcmin \times 8.5\arcmin$ tricolour image R=WISE4, G=WISE3, B=WISE1; b) $2.75\arcmin \times 2.75\arcmin$ Spitzer tricolour image R=irac4, G=irac3, B=irac2; c) NACO $6\arcsec \times 4\arcsec$ image in K. 
       }
         \label{Fighd164794}
   \end{figure*}

Fig.~\ref{Fighd164794} shows the infrared images for HD164794.
The IRAC and WISE images display the circumstellar shell, but it also looks  as if a filament is crossing this star. Is it actually a foreground nebular filament or a jet of this binary system? With the data that we have it is not possible to safely infer the nature of this structure, but this structure appears too large to be a jet.
The MIPS24 image is unfortunately saturated around this star, but it reveals the large complex ambient nebula. In the NACO-K image, the nearby stars previously discussed are visible.

\newpage
\section{Catalogues of stars surrounding HD168625 and the Pistol
Star in NACO images}
\label{catal}
This section gives the catalogue of surrounding objects in the NACO Lp-band field of HD168625 in Table~\ref{tabcathd168625} and in the NACO K-band field of the Pistol
Star in Table~\ref{tabcatPS}. They are provided for archive purposes and as an astrometric reference for any future proper-motion analysis.
The magnitudes in the table are indicated without guarantee for the Pistol
Star due to a possible zeropoint offset of 1 magnitude.
This  results from the lack of usable reference stars with the same instrumental setup as used for the Pistol
Star. Other known stars in the observed field are variable as well, preventing an accurate photometric calibration.

\begin{table}[h!]
\caption{Catalogue of objects surrounding HD168625 in a radius of 18$\arcsec$ as shown in Fig.~\ref{Fighd168625b}.
The NACO Lp magnitude is provided as an indication. }
\centering
\small{
\begin{tabular}{@{\ }l@{\ \ }l@{\ \ }l@{\ \ }l@{\ \ }}
\hline
Star ID & RA(2000) & DEC (2000) & Lp mag.\\
 & h mn s & $\degr$ $\arcmin$ $\arcsec$ & \\
\hline
HD168625a & 18 21 19.49 & -16 22 26.16 & 3.56 $\pm$ 0.00 \\
HD168625b & 18 21 19.46 & -16 22 27.23 & 7.59 $\pm$ 0.00 \\
starD & 18 21 19.57 & -16 22 19.23 & 8.91 $\pm$ 0.00 \\
starL & 18 21 18.58 & -16 22 19.03 & 10.96 $\pm$ 0.03 \\
starM & 18 21 18.71 & -16 22 17.32 & 11.10 $\pm$ 0.03 \\
starC & 18 21 19.22 & -16 22 23.58 & 11.17 $\pm$ 0.03 \\
starJ & 18 21 20.55 & -16 22 29.76 & 11.61 $\pm$ 0.05 \\
starH & 18 21 19.42 & -16 22 13.35 & 12.04 $\pm$ 0.07 \\
starK & 18 21 18.71 & -16 22 20.08 & $>$13 $\pm$ N/A \\
\hline
\end{tabular}
}
\label{tabcathd168625}
\end{table}

\begin{table}[]
\caption{Catalogue of objects surrounding the Pistol
Star within a box of $14\arcsec \times 14\arcsec$ as defined in Fig.~\ref{FigPS}-d.
The NACO K magnitude is provided as an indication. An offset of the zeropoint may exist (magnitudes may have to be corrected by about +1 mag). Entries with K mag values of  99 indicate there are  no reliable measurements. 
Stars with (+) are blended  and their magnitude is not accurate (indicated with :).}
\centering
\tiny{
\begin{tabular}{@{\ }l@{\ \ }l@{\ \ }l@{\ \ }l@{\ \ }}
\hline
Star ID & RA(2000) & DEC (2000) & K mag. \\
 & h mn s & $\degr$ $\arcmin$ $\arcsec$ & \\
\hline
The Pistol
Star & 17 46 15.10 & -28 50 3.58 & 6.75 $\pm$ 0.00 \\
P1  & 17 46 15.09 & -28 50 2.20 & 11.73 $\pm$ 0.01 \\
P2 & 17 46 14.99 & -28 50 3.40 & 11.68 $\pm$ 0.01 \\
P3 & 17 46 15.03 & -28 50 3.90 & 10.85 $\pm$ 0.01 \\
P4 & 17 46 15.07 & -28 50 4.22 & 9.86 $\pm$ 0.00 \\
P5(+P6) & 17 46 15.17 & -28 50 2.45 & 12.44: $\pm$ 0.02 \\
P7 & 17 46 15.24 & -28 50 2.51 & 13.53 $\pm$ 0.04 \\
P8 & 17 46 15.25 & -28 50 2.36 & 13.74 $\pm$ 0.05 \\
P10(+P9+P16) & 17 46 15.01 & -28 50 4.95 & 13.11: $\pm$ 0.03 \\
P11 & 17 46 14.93 & -28 50 4.94 & 14.17 $\pm$ 0.07 \\
P12 & 17 46 15.18 & -28 50 5.08 & 13.01 $\pm$ 0.03 \\
P13 & 17 46 15.17 & -28 50 5.96 & 13.65 $\pm$ 0.04 \\
P14 & 17 46 14.89 & -28 50 3.66 & 13.96 $\pm$ 0.06 \\
P6 & 17 46 15.19 & -28 50 2.24 & 99.00 $\pm$ 99.00 \\
P9  & 17 46 15.02 & -28 50 4.68 & 99.00 $\pm$ 99.00 \\
P15 & 17 46 14.93 & -28 50 1.17 & 99.00 $\pm$ 99.00 \\
P16 & 17 46 14.99 & -28 50 5.11 & 99.00 $\pm$ 99.00 \\
WR102e & 17 46 14.67 & -28 50 0.63 & 9.71 $\pm$ 0.00 \\
(WR102e)b & 17 46 14.70 & -28 50 0.03 & 12.26 $\pm$ 0.02 \\
V4644Sgr & 17 46 14.26 & -28 50 2.61 & 8.46 $\pm$ 0.00 \\
$[FMG99]4$ & 17 46 14.89 & -28 49 53.01 & 10.49 $\pm$ 0.01 \\
$[FMG99]6$ & 17 46 14.25 & -28 49 56.48 & 10.41 $\pm$ 0.00 \\
LHO5 & 17 46 15.82 & -28 49 51.09 & 13.68 $\pm$ 0.05 \\
MLB$\_{PS}$1 & 17 46 15.43 & -28 50 17.62 & 13.08 $\pm$ 0.03 \\
MLB$\_{PS}$2 & 17 46 15.80 & -28 49 50.39 & 13.93 $\pm$ 0.05 \\
MLB$\_{PS}$3 & 17 46 14.44 & -28 49 50.17 & 12.99 $\pm$ 0.02 \\
MLB$\_{PS}$4 & 17 46 14.46 & -28 49 50.31 & 12.77 $\pm$ 0.02 \\
MLB$\_{PS}$5 & 17 46 14.92 & -28 49 50.48 & 14.23 $\pm$ 0.07 \\
MLB$\_{PS}$6 & 17 46 15.11 & -28 49 50.53 & 13.82 $\pm$ 0.05 \\
MLB$\_{PS}$7 & 17 46 14.89 & -28 49 50.62 & 14.50 $\pm$ 0.09 \\
MLB$\_{PS}$8 & 17 46 15.86 & -28 49 50.70 & 17.27 $\pm$ 1.12 \\
MLB$\_{PS}$9 & 17 46 15.25 & -28 49 50.84 & 12.05 $\pm$ 0.01 \\
MLB$\_{PS}$10 & 17 46 15.61 & -28 49 50.91 & 12.27 $\pm$ 0.02 \\
MLB$\_{PS}$11 & 17 46 14.99 & -28 49 50.94 & 11.76 $\pm$ 0.01 \\
MLB$\_{PS}$12 & 17 46 15.80 & -28 49 51.02 & 14.83 $\pm$ 0.12 \\
MLB$\_{PS}$13 & 17 46 15.59 & -28 49 51.14 & 12.14 $\pm$ 0.01 \\
MLB$\_{PS}$14 & 17 46 14.56 & -28 49 51.67 & 13.30 $\pm$ 0.03 \\
MLB$\_{PS}$15 & 17 46 14.83 & -28 49 52.04 & 11.37 $\pm$ 0.01 \\
MLB$\_{PS}$16 & 17 46 14.69 & -28 49 52.35 & 14.84 $\pm$ 0.12 \\
MLB$\_{PS}$17 & 17 46 14.64 & -28 49 52.45 & 13.23 $\pm$ 0.03 \\
MLB$\_{PS}$18 & 17 46 15.58 & -28 49 52.46 & 99.00 $\pm$ 99.00 \\
MLB$\_{PS}$19 & 17 46 14.35 & -28 49 52.53 & 13.63 $\pm$ 0.04 \\
MLB$\_{PS}$20 & 17 46 15.76 & -28 49 52.79 & 12.27 $\pm$ 0.02 \\
MLB$\_{PS}$21 & 17 46 15.23 & -28 49 52.93 & 14.14 $\pm$ 0.07 \\
MLB$\_{PS}$22 & 17 46 15.92 & -28 49 53.02 & 11.74 $\pm$ 0.01 \\
MLB$\_{PS}$23 & 17 46 15.49 & -28 49 53.00 & 15.93 $\pm$ 0.33 \\
MLB$\_{PS}$24 & 17 46 15.12 & -28 49 53.14 & 14.59 $\pm$ 0.10 \\
MLB$\_{PS}$25 & 17 46 14.63 & -28 49 53.30 & 13.46 $\pm$ 0.04 \\
MLB$\_{PS}$26 & 17 46 15.69 & -28 49 53.36 & 12.88 $\pm$ 0.02 \\
MLB$\_{PS}$27 & 17 46 14.67 & -28 49 53.42 & 13.82 $\pm$ 0.05 \\
MLB$\_{PS}$28 & 17 46 15.40 & -28 49 53.51 & 13.37 $\pm$ 0.04 \\
MLB$\_{PS}$29 & 17 46 15.16 & -28 49 53.54 & 13.53 $\pm$ 0.04 \\
MLB$\_{PS}$30 & 17 46 15.70 & -28 49 53.56 & 13.11 $\pm$ 0.03 \\
MLB$\_{PS}$31 & 17 46 15.62 & -28 49 53.69 & 12.20 $\pm$ 0.01 \\
MLB$\_{PS}$32 & 17 46 15.18 & -28 49 53.83 & 13.43 $\pm$ 0.04 \\
MLB$\_{PS}$33 & 17 46 14.34 & -28 49 53.84 & 14.96 $\pm$ 0.14 \\
MLB$\_{PS}$34 & 17 46 15.82 & -28 49 53.88 & 14.09 $\pm$ 0.06 \\
MLB$\_{PS}$35 & 17 46 14.83 & -28 49 54.12 & 13.84 $\pm$ 0.05 \\
MLB$\_{PS}$36 & 17 46 15.40 & -28 49 54.37 & 11.59 $\pm$ 0.01 \\
MLB$\_{PS}$37 & 17 46 15.58 & -28 49 54.31 & 17.13 $\pm$ 0.99 \\
MLB$\_{PS}$38 & 17 46 16.15 & -28 49 54.43 & 13.74 $\pm$ 0.04 \\
MLB$\_{PS}$39 & 17 46 15.14 & -28 49 54.54 & 14.75 $\pm$ 0.11 \\
MLB$\_{PS}$40 & 17 46 14.09 & -28 49 54.58 & 14.06 $\pm$ 0.06 \\
MLB$\_{PS}$41 & 17 46 14.87 & -28 49 54.76 & 11.69 $\pm$ 0.01 \\
MLB$\_{PS}$42 & 17 46 14.62 & -28 49 54.87 & 13.25 $\pm$ 0.03 \\
MLB$\_{PS}$43 & 17 46 16.14 & -28 49 54.91 & 14.10 $\pm$ 0.06 \\
MLB$\_{PS}$44 & 17 46 15.38 & -28 49 55.00 & 12.73 $\pm$ 0.02 \\
MLB$\_{PS}$45 & 17 46 16.10 & -28 49 55.04 & 14.76 $\pm$ 0.11 \\
MLB$\_{PS}$46 & 17 46 14.82 & -28 49 55.07 & 13.87 $\pm$ 0.05 \\
MLB$\_{PS}$47 & 17 46 15.87 & -28 49 55.05 & 16.33 $\pm$ 0.47 \\
MLB$\_{PS}$48 & 17 46 14.05 & -28 49 55.18 & 13.19 $\pm$ 0.03 \\
MLB$\_{PS}$49 & 17 46 15.42 & -28 49 55.21 & 13.00 $\pm$ 0.03 \\
MLB$\_{PS}$50 & 17 46 15.81 & -28 49 55.20 & 13.47 $\pm$ 0.04 \\
\hline
\end{tabular}
}
\label{tabcatPS}
\end{table}
\addtocounter{table}{-1}
\begin{table}[]
\caption{continued}
\centering
\tiny{
\begin{tabular}{@{\ }l@{\ \ }l@{\ \ }l@{\ \ }l@{\ \ }}
\hline
MLB$\_{PS}$51 & 17 46 14.68 & -28 49 55.81 & 12.65 $\pm$ 0.02 \\
MLB$\_{PS}$52 & 17 46 14.48 & -28 49 55.84 & 14.09 $\pm$ 0.06 \\
MLB$\_{PS}$53 & 17 46 14.04 & -28 49 55.86 & 14.56 $\pm$ 0.07 \\
MLB$\_{PS}$54 & 17 46 14.37 & -28 49 55.86 & 13.78 $\pm$ 0.05 \\
MLB$\_{PS}$55 & 17 46 15.67 & -28 49 56.11 & 12.65 $\pm$ 0.02 \\
MLB$\_{PS}$56 & 17 46 14.94 & -28 49 56.18 & 13.42 $\pm$ 0.04 \\
MLB$\_{PS}$57 & 17 46 15.42 & -28 49 56.22 & 14.28 $\pm$ 0.08 \\
MLB$\_{PS}$58 & 17 46 14.34 & -28 49 56.42 & 13.63 $\pm$ 0.04 \\
MLB$\_{PS}$59 & 17 46 14.94 & -28 49 56.46 & 13.54 $\pm$ 0.04 \\
MLB$\_{PS}$60 & 17 46 15.95 & -28 49 56.71 & 14.70 $\pm$ 0.11 \\
MLB$\_{PS}$61 & 17 46 15.16 & -28 49 56.84 & 13.74 $\pm$ 0.05 \\
MLB$\_{PS}$62 & 17 46 14.90 & -28 49 57.38 & 13.40 $\pm$ 0.04 \\
MLB$\_{PS}$63 & 17 46 14.54 & -28 49 57.41 & 14.33 $\pm$ 0.08 \\
MLB$\_{PS}$64 & 17 46 14.99 & -28 49 57.40 & 99.00 $\pm$ 99.00 \\
MLB$\_{PS}$65 & 17 46 16.00 & -28 49 57.47 & 14.81 $\pm$ 0.12 \\
MLB$\_{PS}$66 & 17 46 14.90 & -28 49 57.56 & 13.87 $\pm$ 0.05 \\
MLB$\_{PS}$67 & 17 46 15.55 & -28 49 57.61 & 18.11 $\pm$ 2.42 \\
MLB$\_{PS}$68 & 17 46 14.66 & -28 49 57.82 & 16.33 $\pm$ 0.47 \\
MLB$\_{PS}$69 & 17 46 14.78 & -28 49 57.85 & 16.22 $\pm$ 0.43 \\
MLB$\_{PS}$70 & 17 46 16.13 & -28 49 57.86 & 15.45 $\pm$ 0.20 \\
MLB$\_{PS}$71 & 17 46 14.04 & -28 49 58.22 & 10.89 $\pm$ 0.01 \\
MLB$\_{PS}$72 & 17 46 16.10 & -28 49 58.33 & 14.07 $\pm$ 0.06 \\
MLB$\_{PS}$73 & 17 46 15.03 & -28 49 58.42 & 99.00 $\pm$ 99.00 \\
MLB$\_{PS}$74 & 17 46 14.44 & -28 49 58.48 & 13.33 $\pm$ 0.03 \\
MLB$\_{PS}$75 & 17 46 15.78 & -28 49 58.54 & 13.35 $\pm$ 0.03 \\
MLB$\_{PS}$76 & 17 46 14.11 & -28 49 59.20 & 13.94 $\pm$ 0.06 \\
MLB$\_{PS}$77 & 17 46 15.38 & -28 49 59.21 & 13.76 $\pm$ 0.05 \\
MLB$\_{PS}$78 & 17 46 15.51 & -28 49 59.22 & 13.63 $\pm$ 0.04 \\
MLB$\_{PS}$79 & 17 46 14.74 & -28 49 59.24 & 13.46 $\pm$ 0.04 \\
MLB$\_{PS}$80 & 17 46 14.89 & -28 49 59.33 & 13.60 $\pm$ 0.04 \\
MLB$\_{PS}$81 & 17 46 15.87 & -28 49 59.48 & 14.03 $\pm$ 0.06 \\
MLB$\_{PS}$82 & 17 46 14.06 & -28 49 59.63 & 13.14 $\pm$ 0.03 \\
MLB$\_{PS}$83 & 17 46 14.95 & -28 49 59.66 & 14.54 $\pm$ 0.10 \\
MLB$\_{PS}$84 & 17 46 15.09 & -28 49 59.78 & 13.45 $\pm$ 0.04 \\
MLB$\_{PS}$85 & 17 46 15.82 & -28 49 59.81 & 14.29 $\pm$ 0.08 \\
MLB$\_{PS}$86 & 17 46 14.16 & -28 49 59.96 & 11.27 $\pm$ 0.01 \\
MLB$\_{PS}$87 & 17 46 14.37 & -28 49 59.99 & 14.35 $\pm$ 0.08 \\
MLB$\_{PS}$88 & 17 46 15.77 & -28 50 0.01 & 14.47 $\pm$ 0.09 \\
MLB$\_{PS}$89 & 17 46 15.01 & -28 50 0.09 & 14.23 $\pm$ 0.07 \\
MLB$\_{PS}$90 & 17 46 15.46 & -28 50 0.12 & 14.54 $\pm$ 0.09 \\
MLB$\_{PS}$91 & 17 46 16.06 & -28 50 0.84 & 12.56 $\pm$ 0.02 \\
MLB$\_{PS}$92 & 17 46 15.15 & -28 50 0.94 & 13.99 $\pm$ 0.06 \\
MLB$\_{PS}$93 & 17 46 15.62 & -28 50 0.94 & 11.40 $\pm$ 0.01 \\
MLB$\_{PS}$94 & 17 46 15.63 & -28 50 1.07 & 10.57 $\pm$ 0.01 \\
MLB$\_{PS}$95 & 17 46 15.64 & -28 50 1.22 & 11.26 $\pm$ 0.01 \\
MLB$\_{PS}$96 & 17 46 14.14 & -28 50 1.07 & 13.77 $\pm$ 0.05 \\
MLB$\_{PS}$97 & 17 46 14.39 & -28 50 1.05 & 14.15 $\pm$ 0.07 \\
MLB$\_{PS}$98 & 17 46 15.35 & -28 50 1.10 & 13.97 $\pm$ 0.06 \\
MLB$\_{PS}$99 & 17 46 16.00 & -28 50 1.35 & 13.28 $\pm$ 0.03 \\
MLB$\_{PS}$100 & 17 46 15.48 & -28 50 1.39 & 10.93 $\pm$ 0.01 \\
MLB$\_{PS}$101 & 17 46 14.28 & -28 50 1.44 & 12.88 $\pm$ 0.02 \\
MLB$\_{PS}$102 & 17 46 15.56 & -28 50 1.44 & 12.82 $\pm$ 0.02 \\
MLB$\_{PS}$103 & 17 46 16.08 & -28 50 1.75 & 13.85 $\pm$ 0.05 \\
MLB$\_{PS}$104 & 17 46 15.60 & -28 50 1.84 & 11.33 $\pm$ 0.01 \\
MLB$\_{PS}$105 & 17 46 14.11 & -28 50 1.84 & 12.57 $\pm$ 0.02 \\
MLB$\_{PS}$106 & 17 46 14.15 & -28 50 1.94 & 11.57 $\pm$ 0.01 \\
MLB$\_{PS}$107 & 17 46 15.93 & -28 50 2.64 & 12.66 $\pm$ 0.02 \\
MLB$\_{PS}$108 & 17 46 14.72 & -28 50 2.78 & 14.84 $\pm$ 0.12 \\
MLB$\_{PS}$109 & 17 46 15.63 & -28 50 2.88 & 14.55 $\pm$ 0.10 \\
MLB$\_{PS}$110 & 17 46 14.26 & -28 50 2.99 & 9.47 $\pm$ 0.00 \\
MLB$\_{PS}$111 & 17 46 15.97 & -28 50 3.26 & 15.71 $\pm$ 0.27 \\
MLB$\_{PS}$112 & 17 46 15.04 & -28 50 3.60 & 9.89 $\pm$ 0.00 \\
MLB$\_{PS}$113 & 17 46 14.29 & -28 50 3.69 & 12.66 $\pm$ 0.02 \\
MLB$\_{PS}$114 & 17 46 14.09 & -28 50 3.72 & 13.22 $\pm$ 0.03 \\
MLB$\_{PS}$115 & 17 46 16.05 & -28 50 3.84 & 14.48 $\pm$ 0.09 \\
MLB$\_{PS}$116 & 17 46 14.17 & -28 50 3.96 & 11.01 $\pm$ 0.01 \\
MLB$\_{PS}$117 & 17 46 15.80 & -28 50 4.05 & 13.78 $\pm$ 0.05 \\
MLB$\_{PS}$118 & 17 46 15.50 & -28 50 4.24 & 14.60 $\pm$ 0.10 \\
MLB$\_{PS}$119 & 17 46 14.78 & -28 50 4.44 & 13.95 $\pm$ 0.06 \\
MLB$\_{PS}$120 & 17 46 15.54 & -28 50 4.57 & 14.30 $\pm$ 0.08 \\
MLB$\_{PS}$121 & 17 46 14.69 & -28 50 4.77 & 14.79 $\pm$ 0.12 \\
MLB$\_{PS}$122 & 17 46 14.11 & -28 50 4.99 & 14.50 $\pm$ 0.09 \\
MLB$\_{PS}$123 & 17 46 14.80 & -28 50 5.25 & 13.96 $\pm$ 0.06 \\
MLB$\_{PS}$124 & 17 46 15.48 & -28 50 5.25 & 99.00 $\pm$ 99.00 \\
MLB$\_{PS}$125 & 17 46 15.37 & -28 50 5.39 & 14.89 $\pm$ 0.13 \\
MLB$\_{PS}$126 & 17 46 14.47 & -28 50 5.74 & 13.08 $\pm$ 0.03 \\
MLB$\_{PS}$127 & 17 46 14.26 & -28 50 5.76 & 18.40 $\pm$ 3.17 \\
MLB$\_{PS}$128 & 17 46 14.71 & -28 50 5.79 & 13.31 $\pm$ 0.03 \\
MLB$\_{PS}$129 & 17 46 16.00 & -28 50 5.78 & 14.51 $\pm$ 0.09 \\
MLB$\_{PS}$130 & 17 46 15.62 & -28 50 5.94 & 13.93 $\pm$ 0.06 \\
\hline
\end{tabular}
}
\label{tabcatPS}
\end{table}
\addtocounter{table}{-1}
\begin{table}[]
\caption{continued}
\centering
\tiny{
\begin{tabular}{@{\ }l@{\ \ }l@{\ \ }l@{\ \ }l@{\ \ }}
\hline
MLB$\_{PS}$131 & 17 46 15.38 & -28 50 6.04 & 13.27 $\pm$ 0.03 \\
MLB$\_{PS}$132 & 17 46 14.50 & -28 50 6.14 & 10.89 $\pm$ 0.01 \\
MLB$\_{PS}$133 & 17 46 14.70 & -28 50 6.13 & 12.84 $\pm$ 0.02 \\
MLB$\_{PS}$134 & 17 46 14.39 & -28 50 6.15 & 11.90 $\pm$ 0.01 \\
MLB$\_{PS}$135 & 17 46 15.51 & -28 50 6.19 & 13.90 $\pm$ 0.05 \\
MLB$\_{PS}$136 & 17 46 14.90 & -28 50 6.26 & 14.27 $\pm$ 0.08 \\
MLB$\_{PS}$137 & 17 46 15.46 & -28 50 6.33 & 13.12 $\pm$ 0.03 \\
MLB$\_{PS}$138 & 17 46 15.49 & -28 50 6.44 & 12.99 $\pm$ 0.03 \\
MLB$\_{PS}$139 & 17 46 15.41 & -28 50 6.50 & 14.24 $\pm$ 0.07 \\
MLB$\_{PS}$140 & 17 46 14.04 & -28 50 6.68 & 12.13 $\pm$ 0.01 \\
MLB$\_{PS}$141 & 17 46 15.28 & -28 50 6.66 & 13.95 $\pm$ 0.06 \\
MLB$\_{PS}$142 & 17 46 15.59 & -28 50 6.74 & 12.40 $\pm$ 0.02 \\
MLB$\_{PS}$143 & 17 46 15.81 & -28 50 6.95 & 12.18 $\pm$ 0.01 \\
MLB$\_{PS}$144 & 17 46 14.61 & -28 50 6.96 & 14.45 $\pm$ 0.09 \\
MLB$\_{PS}$145 & 17 46 15.24 & -28 50 7.00 & 12.98 $\pm$ 0.03 \\
MLB$\_{PS}$146 & 17 46 15.56 & -28 50 6.97 & 12.00 $\pm$ 0.01 \\
MLB$\_{PS}$147 & 17 46 15.36 & -28 50 7.12 & 16.08 $\pm$ 0.38 \\
MLB$\_{PS}$148 & 17 46 15.67 & -28 50 7.19 & 12.50 $\pm$ 0.02 \\
MLB$\_{PS}$149 & 17 46 15.43 & -28 50 7.26 & 14.43 $\pm$ 0.09 \\
MLB$\_{PS}$150 & 17 46 15.54 & -28 50 7.36 & 10.93 $\pm$ 0.01 \\
MLB$\_{PS}$151 & 17 46 16.10 & -28 50 7.41 & 11.75 $\pm$ 0.01 \\
MLB$\_{PS}$152 & 17 46 15.56 & -28 50 7.60 & 10.33 $\pm$ 0.00 \\
MLB$\_{PS}$153 & 17 46 14.93 & -28 50 7.60 & 15.73 $\pm$ 0.27 \\
MLB$\_{PS}$154 & 17 46 15.09 & -28 50 7.65 & 13.63 $\pm$ 0.04 \\
MLB$\_{PS}$155 & 17 46 15.47 & -28 50 7.71 & 12.19 $\pm$ 0.01 \\
MLB$\_{PS}$156 & 17 46 15.72 & -28 50 7.89 & 13.89 $\pm$ 0.05 \\
MLB$\_{PS}$157 & 17 46 15.64 & -28 50 8.04 & 13.96 $\pm$ 0.06 \\
MLB$\_{PS}$158 & 17 46 16.06 & -28 50 8.17 & 14.50 $\pm$ 0.09 \\
MLB$\_{PS}$159 & 17 46 15.07 & -28 50 8.26 & 14.35 $\pm$ 0.08 \\
MLB$\_{PS}$160 & 17 46 15.84 & -28 50 8.46 & 14.38 $\pm$ 0.08 \\
MLB$\_{PS}$161 & 17 46 15.54 & -28 50 8.52 & 13.52 $\pm$ 0.04 \\
MLB$\_{PS}$162 & 17 46 14.91 & -28 50 8.79 & 99.00 $\pm$ 99.00 \\
MLB$\_{PS}$163 & 17 46 14.50 & -28 50 9.02 & 14.21 $\pm$ 0.07 \\
MLB$\_{PS}$164 & 17 46 14.33 & -28 50 9.23 & 16.98 $\pm$ 0.86 \\
MLB$\_{PS}$165 & 17 46 15.45 & -28 50 9.35 & 14.10 $\pm$ 0.06 \\
MLB$\_{PS}$166 & 17 46 15.52 & -28 50 9.34 & 14.08 $\pm$ 0.06 \\
MLB$\_{PS}$167 & 17 46 16.11 & -28 50 9.56 & 14.92 $\pm$ 0.13 \\
MLB$\_{PS}$168 & 17 46 15.80 & -28 50 9.71 & 14.78 $\pm$ 0.12 \\
MLB$\_{PS}$169 & 17 46 15.20 & -28 50 10.10 & 14.44 $\pm$ 0.09 \\
MLB$\_{PS}$170 & 17 46 14.39 & -28 50 10.13 & 14.65 $\pm$ 0.10 \\
MLB$\_{PS}$171 & 17 46 15.31 & -28 50 10.26 & 13.10 $\pm$ 0.03 \\
MLB$\_{PS}$172 & 17 46 16.07 & -28 50 10.67 & 14.18 $\pm$ 0.07 \\
MLB$\_{PS}$173 & 17 46 15.37 & -28 50 10.70 & 13.89 $\pm$ 0.05 \\
MLB$\_{PS}$174 & 17 46 15.57 & -28 50 10.76 & 14.49 $\pm$ 0.09 \\
MLB$\_{PS}$175 & 17 46 14.52 & -28 50 10.78 & 14.31 $\pm$ 0.08 \\
MLB$\_{PS}$176 & 17 46 14.35 & -28 50 10.92 & 14.52 $\pm$ 0.09 \\
MLB$\_{PS}$177 & 17 46 15.07 & -28 50 11.00 & 13.91 $\pm$ 0.06 \\
MLB$\_{PS}$178 & 17 46 15.74 & -28 50 11.19 & 12.18 $\pm$ 0.01 \\
MLB$\_{PS}$179 & 17 46 15.50 & -28 50 11.29 & 14.31 $\pm$ 0.08 \\
MLB$\_{PS}$180 & 17 46 14.63 & -28 50 11.38 & 10.64 $\pm$ 0.01 \\
MLB$\_{PS}$181 & 17 46 14.70 & -28 50 11.46 & 13.31 $\pm$ 0.03 \\
MLB$\_{PS}$182 & 17 46 15.03 & -28 50 11.95 & 11.92 $\pm$ 0.01 \\
MLB$\_{PS}$183 & 17 46 15.00 & -28 50 12.05 & 11.34 $\pm$ 0.01 \\
MLB$\_{PS}$184 & 17 46 14.78 & -28 50 12.02 & 15.09 $\pm$ 0.16 \\
MLB$\_{PS}$185 & 17 46 15.50 & -28 50 12.17 & 14.53 $\pm$ 0.09 \\
MLB$\_{PS}$186 & 17 46 15.79 & -28 50 12.25 & 14.53 $\pm$ 0.09 \\
MLB$\_{PS}$187 & 17 46 15.21 & -28 50 12.29 & 13.88 $\pm$ 0.05 \\
MLB$\_{PS}$188 & 17 46 16.14 & -28 50 12.59 & 10.72 $\pm$ 0.01 \\
MLB$\_{PS}$189 & 17 46 14.38 & -28 50 12.82 & 11.19 $\pm$ 0.01 \\
MLB$\_{PS}$190 & 17 46 15.34 & -28 50 12.79 & 13.85 $\pm$ 0.05 \\
MLB$\_{PS}$191 & 17 46 15.33 & -28 50 12.84 & 13.92 $\pm$ 0.06 \\
MLB$\_{PS}$192 & 17 46 15.64 & -28 50 12.98 & 12.86 $\pm$ 0.02 \\
MLB$\_{PS}$193 & 17 46 15.02 & -28 50 13.18 & 15.80 $\pm$ 0.29 \\
MLB$\_{PS}$194 & 17 46 14.68 & -28 50 13.71 & 14.32 $\pm$ 0.08 \\
MLB$\_{PS}$195 & 17 46 15.52 & -28 50 13.93 & 13.05 $\pm$ 0.03 \\
MLB$\_{PS}$196 & 17 46 15.94 & -28 50 14.07 & 12.43 $\pm$ 0.02 \\
MLB$\_{PS}$197 & 17 46 15.47 & -28 50 14.15 & 10.63 $\pm$ 0.01 \\
MLB$\_{PS}$198 & 17 46 15.33 & -28 50 14.24 & 13.86 $\pm$ 0.05 \\
MLB$\_{PS}$199 & 17 46 15.62 & -28 50 14.27 & 12.51 $\pm$ 0.02 \\
MLB$\_{PS}$200 & 17 46 15.36 & -28 50 14.39 & 13.46 $\pm$ 0.04 \\
MLB$\_{PS}$201 & 17 46 15.44 & -28 50 14.40 & 11.45 $\pm$ 0.01 \\
MLB$\_{PS}$202 & 17 46 14.60 & -28 50 14.42 & 14.21 $\pm$ 0.07 \\
MLB$\_{PS}$203 & 17 46 14.69 & -28 50 14.52 & 14.01 $\pm$ 0.06 \\
MLB$\_{PS}$204 & 17 46 15.42 & -28 50 14.58 & 12.63 $\pm$ 0.02 \\
MLB$\_{PS}$205 & 17 46 15.13 & -28 50 14.77 & 14.17 $\pm$ 0.07 \\
MLB$\_{PS}$206 & 17 46 15.31 & -28 50 15.20 & 14.15 $\pm$ 0.07 \\
MLB$\_{PS}$207 & 17 46 15.67 & -28 50 15.40 & 12.74 $\pm$ 0.02 \\
MLB$\_{PS}$208 & 17 46 15.95 & -28 50 15.68 & 13.87 $\pm$ 0.05 \\
MLB$\_{PS}$209 & 17 46 15.60 & -28 50 15.85 & 12.85 $\pm$ 0.02 \\
MLB$\_{PS}$210 & 17 46 15.56 & -28 50 16.53 & 13.37 $\pm$ 0.04 \\
MLB$\_{PS}$211 & 17 46 15.66 & -28 50 16.50 & 14.85 $\pm$ 0.12 \\
MLB$\_{PS}$212 & 17 46 14.90 & -28 50 16.82 & 13.96 $\pm$ 0.06 \\
MLB$\_{PS}$213 & 17 46 14.86 & -28 50 16.96 & 15.04 $\pm$ 0.15 \\
MLB$\_{PS}$214 & 17 46 15.84 & -28 50 17.05 & 13.47 $\pm$ 0.04 \\
MLB$\_{PS}$215 & 17 46 14.30 & -28 50 17.26 & 14.08 $\pm$ 0.06 \\
\hline
\end{tabular}
}
\label{tabcatPS}
\end{table}

\end{appendix} 

\end{document}